# Antisymmetry rules of response properties in certain chemical spaces

Takafumi Shiraogawa,[1,2,3,a)] Simon León Krug,[4] Masahiro Ehara,[1,2,3,b)] and O. Anatole von Lilienfeld[4,5,6,7,8,9,10,c)]

[1]*Institute for Molecular Science, National Institutes of Natural Sciences, 38 Nishigonaka, Myodaiji, Okazaki 444-8585, Japan*

[2]*Research Center for Computational Science, National Institutes of Natural Sciences, 38 Nishigonaka, Myodaiji, Okazaki 444-8585, Japan*

[3]*The Graduate University for Advanced Studies, 38 Nishigonaka, Myodaiji, Okazaki 444-8585, Japan*

[4]*Machine Learning Group, Technische Universität Berlin, 10587 Berlin, Germany*

[5]*Chemical Physics Theory Group, Department of Chemistry, University of Toronto, St. George Campus, Toronto, M5S3H6 Ontario, Canada*

[6]*Department of Materials Science and Engineering, University of Toronto, St. George Campus, Toronto, M5S 3E4 Ontario, Canada*

[7]*Vector Institute for Artificial Intelligence, Toronto, M5S 1M1 Ontario, Canada*

[8]*Berlin Institute for the Foundations of Learning and Data, 10587 Berlin, Germany*

[9]*Department of Physics, University of Toronto, St. George Campus, Toronto, M5S 1A7 Ontario, Canada*

[10]*Acceleration Consortium, University of Toronto, Toronto, M5R 0A3 Ontario, Canada*

Correspondence to: a) shiraogawa@ims.ac.jp; b) ehara@ims.ac.jp; c) anatole.vonlilienfeld@utoronto.ca

## Abstract

Understanding chemical compound space (CCS), a set of molecules and materials, is crucial for the rational discovery of molecules and materials. Concepts of symmetry have recently been introduced into CCS to account for near degeneracies and differences in electronic energies between iso-electronic materials. In this work, we present approximate relationships of response properties based on a first-principles view of CCS. They have been derived from perturbation theory and antisymmetry considerations involving nuclear charges. These rules allow approximate predictions of relative response properties of pairs of distinct compounds with opposite nuclear charge variations from a highly symmetric reference material, without the need for experiments or quantum chemical calculations of each compound. We numerically and statistically verified these rules for electric and magnetic response properties (electric dipole moment, polarizabilities, hyperpolarizabilities, and magnetizabilities) among charge-neutral and iso-electronic boron nitride-doped polycyclic aromatic hydrocarbon derivatives of naphthalene, anthracene, and pyrene. Our analysis indicates that, despite their simplicity, antisymmetry rule-based predictions are remarkably accurate, enabling dimensionality reduction of CCS. The rules predict the electric response properties more accurately than the magnetizabilities. The electric response properties in alchemical perturbation density functional theory were investigated to clarify the origin of this predictive power.

# I. INTRODUCTION

The design of compounds can be regarded as an exploration of chemical compound space (CCS), a set of molecules and materials. The rational exploration of CCS accelerates the development of materials with desirable properties. However, the combinatorial scaling of CCS due to various possible combinations of atom types and positions makes this challenging.[1] For example, experimentally or computationally enumerating the properties of all compounds, ignoring underlying relationships among materials in CCS, is conceptually straightforward but often infeasible in practice. Uncovering fundamental principles of physics and chemistry that govern material properties can narrow down and guide CCS. Therefore, deepening our comprehensive understanding of CCS is of fundamental importance in chemistry and materials science.

In quantum alchemy, relationships between different materials are described by the continuous interpolation of Hamiltonians, which has been used to investigate changes in properties due to interconversions of materials.[2-41] When the Hamiltonian is defined by a set of nuclear charges, nuclear coordinates, and the number of electrons, quantum alchemy is consistent with the first-principles view of CCS.[42] By treating these changes as perturbations, alchemical perturbation density functional theory (APDFT)[43] can predict the properties of other materials based on the electronic structure of a reference material. Applications of APDFT involve reaction energies,[25, 43-45] catalysis,[30, 36, 39] response properties,[41, 43] and spin-excited states.[46] Recently, comprehensive perspectives of CCS based on quantum alchemy have been developed, involving energy partitioning,[47, 48] chemical bonding,[49-51] and symmetry.[49, 52] The alchemical integral transform,[53, 54] derived from APDFT, has discovered an approximate quadratic dependence of relative energies between iso-electronic atoms on their nuclear charges.[55] These concepts are universally applicable to various materials owing to their quantum mechanics foundations, within the scope of their approximations.

The concept of alchemical chirality introduces the notion of symmetry into CCS. The alchemical chirality shows that the electronic energies of two disparate iso-electronic materials, called alchemical enantiomers, are nearly identical.[49] Analogous to the established enantiomers of chiral molecules, alchemical enantiomers are defined as non-superimposable materials in a four-dimensional space spanned by nuclear coordinates and charges. Moreover, alchemical enantiomers are compounds sharing the same geometry in which nuclear charge differences from a reflection plane (reference material) are in mirror symmetry relationships. Within each enantiomer, these differences cancel out in pairs of atom sites in equivalent reference's chemical environments (Fig. 1(a)). The reference corresponds to a maximum in electronic energy among some adjacent alchemical enantiomers, *i.e.*, its first-order alchemical derivative vanishes.

When the symmetry requirement of alchemical chirality is lifted, mirror images are called alchemical diastereomers.[52] Fig. 1(b) shows examples of alchemical diastereomers of boron nitride (BN)-doped naphthalene derivatives and cubic-octahedral 79-atom nanoclusters. The transmuted carbons at the $\alpha$ and $\beta$ positions in reference naphthalene have different chemical environments. The transmutation of atoms happens on the (111) facet and edge in the reference nanocluster, which exhibit distinct chemical characteristics. Electronic energies of alchemical diastereomers are no longer degenerate, and the first-order alchemical derivative accurately describes their relative energies.[52] From the perspective of perturbation theory, alchemical enantiomers and diastereomers can be viewed as pairs of materials connected by antisymmetric perturbations on a reference material.[49, 52] These pairs are not only geometric isomers but also materials with different chemical compositions.

In this work, we introduce a concept of symmetry in response properties to external fields in iso-electronic CCS to reduce its formal dimensionality. Our antisymmetry rules propose simple relationships between alchemical enantiomers and diastereomers, defined from a highly symmetric reference system (Fig. 1(a)). The antisymmetry rules have been derived from perturbation theory and antisymmetry considerations involving nuclear charges, which are transferable across various CCSs. We numerically investigated the applicability and limitations of the antisymmetry rules. We examined various response properties, including electric dipole moments, polarizabilities, hyperpolarizabilities, and magnetizabilities (magnetic susceptibilities) of paramagnetic molecules. Our numerical analysis was performed on approximately 3,100 boron nitride (BN)-doped polycyclic aromatic hydrocarbon (PAH) derivatives in the flameworks of naphthalene, anthracene, and pyrene.[56] BN-doped PAH derivatives are promising candidates for applications in energy storage, energy conversion, catalysis, sensing, and gas storage.[57-59] We also examined the electric response properties in the framework of APDFT, and the numerical results were compared with those of the antisymmetry rule-based predictions.

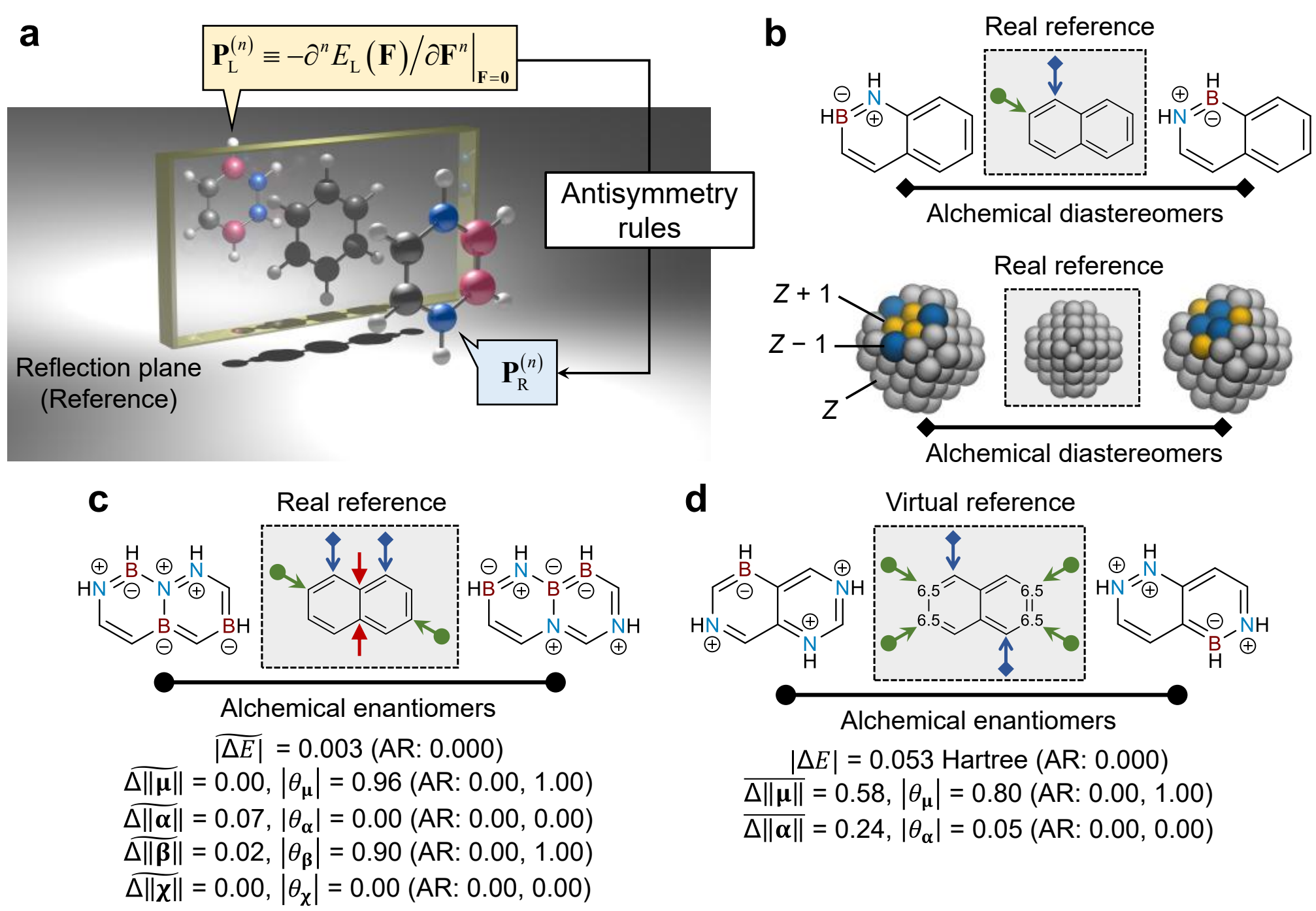

**FIG 1.** Illustrations of the antisymmetry rules (ARs) of response properties of materials. (a) Antisymmetry rules for a pair of alchemical enantiomers or diastereomers (L and R) of BN-doped benzene derivatives with a reflection plane (reference benzene) in CCS. Carbon, gray; boron, pink; nitrogen, blue. (b) Alchemical diastereomers of BN-doped naphthalene derivatives and cuboctahedron 79-atom nanoclusters. $Z$ denotes the nuclear charge of an atom. (c) Antisymmetry rules for a pair of alchemical enantiomers of BN-doped naphthalene derivatives. The absolute relative magnitude of the property $\mathbf{P}$, $\widetilde{\Delta\|\mathbf{P}\|}$, is divided by the corresponding median for randomly selected pairs (see Figs. S5 and 2). $|\theta_{\mathbf{P}}|$ is the absolute relative angle (in $\pi$ radians) between the response properties. (d) Virtual reference molecule-based antisymmetry rules for a pair of alchemical enantiomers of BN-doped naphthalene derivatives. The geometry was fixed to the locally stable structure of naphthalene. The nuclear charges of the reference's virtual atoms are 6.5. $\overline{\Delta\|\mathbf{P}\|}$ denotes the absolute relative property divided by the average of the respective absolute values.

## II. THEORY

Our introduced antisymmetry rules are simple relationships between response properties ($\mathbf{P}_{\mathrm{L}}^{(n)} \equiv -\partial^n E_{\mathrm{L}}(\mathbf{F})/\partial \mathbf{F}^n\big|_{\mathbf{F}=\mathbf{0}}$ and $\mathbf{P}_{\mathrm{R}}^{(n)}$) of pairs (L and R) of compounds of alchemical enantiomers and diastereomers, defined by charge-neutral mutations of atoms of a highly symmetric reference material. $\mathbf{P}^{(n)}$ is the $n$th-order response property. $E$ is the total energy. $\mathbf{F}$ is the amplitude of an external field. L and R are the iso-electronic compounds. In the following, the antisymmetry rules are derived from APDFT and symmetry considerations.

In APDFT, the electronic Hamiltonian is represented by continuously interpolating those of reference and target materials ($\hat{H}_{\text{Ref}}$ and $\hat{H}_{\text{target}}$) as $\hat{H}(\lambda) = \hat{H}_{\text{Ref}} + \hat{H}'(\lambda) = \hat{H}_{\text{Ref}} + \lambda\left(\hat{H}_{\text{target}} - \hat{H}_{\text{Ref}}\right)$, where $\lambda$ ($0 \le \lambda \le 1, \lambda \in \mathbb{R}$) is a one-dimensional coupling parameter.[43] When the targets are L and R, $\hat{H}_{\text{target}} - \hat{H}_{\text{Ref}}$ is represented as $\hat{H}_{\text{L}} - \hat{H}_{\text{Ref}} = \Delta v(\mathbf{r})$ and $\hat{H}_{\text{R}} - \hat{H}_{\text{Ref}} = -\Delta v(\mathbf{r})$ with the difference in external potentials, $\Delta v(\mathbf{r}) \equiv -\sum_I \Delta Z_I / |\mathbf{r} - \mathbf{R}_I|$, and perturbation $\hat{H}'(\lambda)$ is antisymmetric for L and R.[49, 52] $\Delta Z_I$ is the difference in the $I$th nuclear charge between the reference compound and L. The geometries of L and R are assumed to be fixed to that of the reference compound. $\hat{H}(\lambda)$ is rewritten as $\hat{H}(\lambda) = \hat{H}_{\text{Ref}} + \lambda \Delta v(\mathbf{r})$ with $-1 \le \lambda \le 1$. Clearly, $\hat{H}(\lambda = 0) = \hat{H}_{\text{Ref}}$, $\hat{H}(\lambda = +1) = \hat{H}_{\text{L}}$, and $\hat{H}(\lambda = -1) = \hat{H}_{\text{R}}$. The Taylor series expansions with this $\hat{H}(\lambda)$ around the reference give electronic energies of L and R:[52]

$$E_{\text{L}} = E(\lambda = +1) = E_{\text{Ref}} + \frac{\partial E_{\text{Ref}}}{\partial \lambda} + \frac{1}{2}\frac{\partial^2 E_{\text{Ref}}}{\partial \lambda^2} + \ldots \quad (1)$$

$$E_{\text{R}} = E(\lambda = -1) = E_{\text{Ref}} - \frac{\partial E_{\text{Ref}}}{\partial \lambda} + \frac{1}{2}\frac{\partial^2 E_{\text{Ref}}}{\partial \lambda^2} - \ldots \quad (2)$$

where $E_{\text{Ref}} = E(\lambda = 0)$ denotes the electronic energy of the reference compound. The convergence of the perturbation expansion has been confirmed in several systems.[60] Electronic response properties to a static and uniform external field are expressed as derivatives of the electronic energy with respect to the external field $\mathbf{F}$:

$$\begin{aligned}\mathbf{P}_{\text{L}}^{(n)} = &-\left.\frac{\partial^n E_{\text{Ref}}(\mathbf{F})}{\partial \mathbf{F}^n}\right|_{\mathbf{F}=\mathbf{0}} - \left.\frac{\partial^{1+n} E_{\text{Ref}}(\mathbf{F})}{\partial \lambda \partial \mathbf{F}^n}\right|_{\mathbf{F}=\mathbf{0}} \\ &- \frac{1}{2}\left.\frac{\partial^{2+n} E_{\text{Ref}}(\mathbf{F})}{\partial \lambda^2 \partial \mathbf{F}^n}\right|_{\mathbf{F}=\mathbf{0}} - \ldots\end{aligned} \quad (3)$$

$$\begin{aligned}\mathbf{P}_{\text{R}}^{(n)} = &-\left.\frac{\partial^n E_{\text{Ref}}(\mathbf{F})}{\partial \mathbf{F}^n}\right|_{\mathbf{F}=\mathbf{0}} + \left.\frac{\partial^{1+n} E_{\text{Ref}}(\mathbf{F})}{\partial \lambda \partial \mathbf{F}^n}\right|_{\mathbf{F}=\mathbf{0}} \\ &- \frac{1}{2}\left.\frac{\partial^{2+n} E_{\text{Ref}}(\mathbf{F})}{\partial \lambda^2 \partial \mathbf{F}^n}\right|_{\mathbf{F}=\mathbf{0}} + \ldots\end{aligned} \quad (4)$$

Within the second-order perturbation expansion, subtracting $\mathbf{P}_{\text{R}}^{(n)}$ from $\mathbf{P}_{\text{L}}^{(n)}$ yields

$$\mathbf{P}_{\text{L}}^{(n)} - \mathbf{P}_{\text{R}}^{(n)} \approx -2\left.\frac{\partial^{1+n} E_{\text{Ref}}(\mathbf{F})}{\partial \lambda \partial \mathbf{F}^n}\right|_{\mathbf{F}=\mathbf{0}} \quad (5)$$

where the even-order terms vanish. When the energy derivatives can be calculated, this equation itself is useful for estimating relative properties up to the second-order perturbation expansion. For a variationally optimized wavefunction, according to the Hellmann–Feynman theorem,

$$\mathbf{P}_{\text{L}}^{(n)} - \mathbf{P}_{\text{R}}^{(n)} \approx -2\int d\mathbf{r}\, \Delta v(\mathbf{r}) \left.\frac{\partial^n \rho_{\text{Ref}}(\mathbf{r}, \mathbf{F})}{\partial \mathbf{F}^n}\right|_{\mathbf{F}=\mathbf{0}} \quad (6)$$

where $\rho_{\text{Ref}}(\mathbf{r}, \mathbf{F})$ is the electron density of the reference compound, $\rho_{\text{Ref}}(\mathbf{r})$, in the presence of an external field.

We derive antisymmetry rules for odd-order response properties from the above equations and symmetry considerations. We can choose a reference compound with spatial symmetry that makes the response properties vanish (*i.e.*, $-\partial^n E_{\text{Ref}}(\mathbf{F})/\partial \mathbf{F}^n|_{\mathbf{F}=\mathbf{0}} = \mathbf{0}$). For example, molecules with a center of symmetry satisfy this condition. Substituting this equation into Eqs. (3) and (4) gives

$$\mathbf{P}_{\text{L}}^{(n)} = -\left.\frac{\partial^{1+n} E_{\text{Ref}}(\mathbf{F})}{\partial \lambda \partial \mathbf{F}^n}\right|_{\mathbf{F}=\mathbf{0}} - \frac{1}{2}\left.\frac{\partial^{2+n} E_{\text{Ref}}(\mathbf{F})}{\partial \lambda^2 \partial \mathbf{F}^n}\right|_{\mathbf{F}=\mathbf{0}} - \ldots \quad (7)$$

$$\mathbf{P}_{\text{R}}^{(n)} = \left.\frac{\partial^{1+n} E_{\text{Ref}}(\mathbf{F})}{\partial \lambda \partial \mathbf{F}^n}\right|_{\mathbf{F}=\mathbf{0}} - \frac{1}{2}\left.\frac{\partial^{2+n} E_{\text{Ref}}(\mathbf{F})}{\partial \lambda^2 \partial \mathbf{F}^n}\right|_{\mathbf{F}=\mathbf{0}} + \ldots \quad (8)$$

Within the first-order perturbation expansion, summing $\mathbf{P}_{\text{L}}^{(n)}$ and $\mathbf{P}_{\text{R}}^{(n)}$ results in canceling the odd-order terms:

$$\mathbf{P}_{\text{L}}^{(n)} \approx -\mathbf{P}_{\text{R}}^{(n)} \quad (9)$$

This antisymmetry rule shows that magnitudes of odd-order response properties are approximately the same, yet their orientations are the opposite. Within the perturbation approximation, it is valid for both the alchemical enantiomers and diastereomers. Eq. (9) can be directly derived by differentiating the electronic energy relationship[52] with respect to $\mathbf{F}$. For paramagnetic molecules, $-\partial^n E(\mathbf{B})/\partial \mathbf{B}^n|_{\mathbf{B}=\mathbf{0}} = \mathbf{0}$ $(n = 1, 3, 5, \ldots)$ with magnetic field $\mathbf{B}$.[61] We note that including the term of nuclear

charges in the electric dipole moment does not alter the relation in Eq. (9).

By considering symmetry in Eq. (6), antisymmetry rules for even-order response properties are derived. Unlike the odd-order case, it is usually not possible to choose the reference system so that the zeroth-order term vanishes. If pairs of distinct transmuting atoms in the reference system have identical chemical environments for $\partial^n \rho_{\mathrm{Ref}}(\mathbf{r},\mathbf{F})/\partial \mathbf{F}^n\big|_{\mathbf{F}=\mathbf{0}}$, the integral in Eq. (6) vanishes since $\sum_I \Delta Z_I = 0$:

$$\mathbf{P}_{\mathrm{L}}^{(n)} - \mathbf{P}_{\mathrm{R}}^{(n)} \approx -2\int d\mathbf{r}\,\Delta v(\mathbf{r}) \left.\frac{\partial^n \rho_{\mathrm{Ref}}(\mathbf{r},\mathbf{F})}{\partial \mathbf{F}^n}\right|_{\mathbf{F}=\mathbf{0}} = \mathbf{0} \tag{10}$$

When this equation is satisfied, $\mathbf{P}_{\mathrm{L}}^{(n)}$ and $\mathbf{P}_{\mathrm{R}}^{(n)}$ are approximately the same. When $n = 0$, this relation corresponds to the approximate degeneracy of electronic energies of alchemical enantiomers.[49] Identical environments are mapped onto each other by a spatial symmetry operation. For alchemical diastereomers, however, not all pairs of transmuting atoms are symmetrically equivalent, and Eq. (10) is no longer satisfied. Nevertheless, some alchemical diastereomers exhibit near degeneracy in electronic energies and are considered approximate alchemical enantiomers.[49]

For even-order response properties ($n > 0$) of alchemical enantiomers, it is necessary to additionally consider the spatial symmetry of $\partial^n \rho_{\mathrm{Ref}}(\mathbf{r},\mathbf{F})/\partial \mathbf{F}^n\big|_{\mathbf{F}=\mathbf{0}}$. Eq. (10) is satisfied for pairs of transmuting atoms if $\partial^n \rho_{\mathrm{Ref}}(\mathbf{r},\mathbf{F})/\partial \mathbf{F}^n\big|_{\mathbf{F}=\mathbf{0}}$ is symmetric under a symmetry operation. We have shown the symmetry of the chemical environments mapped onto each other through inversion and reflection (supplementary material). We assume that the reference compound is in the principal axis coordinate. For inversion, $\partial^n \rho_{\mathrm{Ref}}(\mathbf{r},\mathbf{F})/\partial \mathbf{F}^n\big|_{\mathbf{F}=\mathbf{0}}$ is symmetric. For reflection, $\partial^n \rho_{\mathrm{Ref}}(\mathbf{r},\mathbf{F})/\partial \mathbf{F}^n\big|_{\mathbf{F}=\mathbf{0}}$ contains both symmetric and antisymmetric elements, and therefore Eq. (10) may not always hold. For example, the spatial reflection of the diagonal and off-diagonal elements $\partial^2 \rho_{\mathrm{Ref}}(\mathbf{r},\mathbf{F})/\partial F_i \partial F_j\big|_{\mathbf{F}=\mathbf{0}}$ for second-order response properties maintains and changes their sign along the reflection direction, respectively. However, Eq. (9) can be applied to the off-diagonal elements because the corresponding components of the zeroth-order term are zero. For rotation, the element purely along the rotation axis, $\partial^2 \rho_{\mathrm{Ref}}(\mathbf{r},\mathbf{F})/\partial {F_k}^2\big|_{\mathbf{F}=\mathbf{0}}$, is symmetric. We will consider the other two diagonal elements of second-order response properties. Twofold rotational symmetry leads to Eq. (10). We have obtained the rule for threefold or higher rotational symmetry within the first-order approximation. The diagonal elements of the zeroth-order term in Eqs. (3) and (4) are identical (*i.e.*, $\partial^2 E_{\mathrm{Ref}}(\mathbf{F})/\partial F_i^2\big|_{\mathbf{F}=\mathbf{0}} = \partial^2 E_{\mathrm{Ref}}(\mathbf{F})/\partial {F_j}^2\big|_{\mathbf{F}=\mathbf{0}}$ ). For the first-order term, because $\partial^3 E_{\mathrm{Ref}}(\mathbf{F})/\partial Z_I \partial F_i^2\big|_{\mathbf{F}=\mathbf{0}} + \partial^3 E_{\mathrm{Ref}}(\mathbf{F})/\partial Z_I \partial {F_j}^2\big|_{\mathbf{F}=\mathbf{0}}$ is invariant under rotation, we obtain $\partial^3 E_{\mathrm{Ref}}(\mathbf{F})/\partial\lambda \partial F_i^2\big|_{\mathbf{F}=\mathbf{0}} = -\partial^3 E_{\mathrm{Ref}}(\mathbf{F})/\partial\lambda \partial {F_j}^2\big|_{\mathbf{F}=\mathbf{0}}$. Using those relations, the difference between the diagonal elements is expressed from Eq. (3) as

$$P_{\mathrm{L},ii}^{(2)} - P_{\mathrm{L},jj}^{(2)} \approx -2 \left.\frac{\partial^3 E_{\mathrm{Ref}}(\mathbf{F})}{\partial\lambda \partial F_i^2}\right|_{\mathbf{F}=\mathbf{0}} = 2 \left.\frac{\partial^3 E_{\mathrm{Ref}}(\mathbf{F})}{\partial\lambda \partial {F_j}^2}\right|_{\mathbf{F}=\mathbf{0}} \tag{11}$$

Substituting this equation into Eq. (5) results in

$$P_{\mathrm{L},ii}^{(2)} \approx P_{\mathrm{R},jj}^{(2)} \tag{12}$$

The off-diagonal elements follow the same rule as odd-order response properties (Eq. (9)) since those of the zeroth-order term are zero. Our numerical calculations of electric polarizabilities show that the off-diagonal elements of the first-order term are small, indicating the importance of higher-order perturbation contributions. Therefore, Eq. (10) was applied to the off-diagonal elements instead of Eq. (9).

For even-order response properties of alchemical diastereomers, Eq. (10) is not strictly exact due to the lack of spatial symmetry ensuring pairs of equivalent chemical environments for all transmuting atoms. However, we applied Eq. (10) as an approximation to alchemical diastereomers.

The resulting antisymmetry rules enable a simple estimation of $\mathbf{P}_{\mathrm{R}}^{(n)}$ from $\mathbf{P}_{\mathrm{L}}^{(n)}$ based on Eqs. (9) and (10) within the truncated perturbation expansion (Fig. 1(a)). Notably, we can choose unstable or unrealistic reference structures, as the reference's electronic structure and physical properties are not required. The antisymmetry rules are predicated on the assumption that CCS is predominantly smooth. In general, higher-order perturbation terms in APDFT are sources of error. Furthermore, geometry relaxation typically lifts the antisymmetry rules.

Our current examination of the antisymmetry rules does not cover all spatial symmetries and higher even-order response properties. A more systematic study of the symmetry of chemical environments could expand the applicability of the antisymmetry rules.

From the viewpoint of conceptual DFT (CDFT), for instance, the simple relationship between electronegativity (or chemical potential) and electric dipole moment for diatomic molecules was derived from the electronegativity equalization principle.[62] Typically, the electronegativities of isolated atoms are used. For more complex systems, the bond dipole model,[63] where the molecular dipole moment is expressed as a vector sum of the bond dipole moments in the molecule, is often used. Here, we provide an example of the CDFT dipole moment for comparison with the antisymmetry rule. For a neutral diatomic molecule AB, the partial charge on atom A can be represented as[62]

$$q_{\mathrm{A}} = \frac{\chi_{\mathrm{B}} - \chi_{\mathrm{A}}}{2\left(\eta_{\mathrm{A}} + \eta_{\mathrm{B}}\right)} \tag{13}$$

where $\chi_{\mathrm{A}}$ and $\eta_{\mathrm{A}}$ are the electronegativity and hardness of a neutral atom A, respectively. Using this equation with $q_{\mathrm{A}} + q_{\mathrm{B}} = 0$, the dipole moment for AB is expressed as

$$\boldsymbol{\mu}_{\mathrm{AB}} = q_{\mathrm{A}}\mathbf{R}_{\mathrm{A}} + q_{\mathrm{B}}\mathbf{R}_{\mathrm{B}} = \frac{\chi_{\mathrm{B}} - \chi_{\mathrm{A}}}{2\left(\eta_{\mathrm{A}} + \eta_{\mathrm{B}}\right)}\left(\mathbf{R}_{\mathrm{A}} - \mathbf{R}_{\mathrm{B}}\right) \tag{14}$$

where $\mathbf{R}_{\mathrm{A}}$ denotes the coordinates of A. Combining with the bond dipole model, the molecular dipole moment can be represented as a sum of $\boldsymbol{\mu}_{\mathrm{AB}}$ over all the bonds:

$$\boldsymbol{\mu} = \sum_I \boldsymbol{\mu}_{\mathrm{AB},I} \tag{15}$$

where *I* runs over all the bonds. Using this equation, the vector sum of the dipole moments for the antisymmetric pair L and R yields

$$\boldsymbol{\mu}_{\mathrm{L}} + \boldsymbol{\mu}_{\mathrm{R}} = \frac{1}{2}\sum_I \left( \frac{\chi^{\mathrm{L}}_{\mathrm{B},I} - \chi^{\mathrm{L}}_{\mathrm{A},I}}{\eta^{\mathrm{L}}_{\mathrm{A},I} + \eta^{\mathrm{L}}_{\mathrm{B},I}} + \frac{\chi^{\mathrm{R}}_{\mathrm{B},I} - \chi^{\mathrm{R}}_{\mathrm{A},I}}{\eta^{\mathrm{R}}_{\mathrm{A},I} + \eta^{\mathrm{R}}_{\mathrm{B},I}} \right)\left(\mathbf{R}^{\mathrm{L}}_{\mathrm{A},I} - \mathbf{R}^{\mathrm{L}}_{\mathrm{B},I}\right) \tag{16}$$

where we assume identical bonding structures for L and R. This sum does not generally vanish, clearly showing the difference between the CDFT representation and the antisymmetry rule (Eq. (9)). Moreover, the antisymmetry rules have been derived without approximations on atoms and chemical bonds in molecules. Considering the potential connection between CDFT and quantum alchemy more deeply would be an interesting case for future study.

Here, we compare the computational costs of the antisymmetry rules, APDFT, and standard quantum chemistry methods. While the standard quantum chemistry approach requires calculations for each compound in CCS, APDFT relies on a single explicit calculation on a reference to predict properties of isoelectronic and similar compounds. The antisymmetry rules, while having a more limited scope of application to alchemical enantiomers and diastereomers, can predict (relative) response properties without any quantum chemistry calculations or experiments.

In this work, electric dipole moment (**μ**), electric polarizabilities (**α**), electric hyperpolarizabilities (**β**), and magnetizabilities (**χ**), expressed by the following equations, are considered as response properties:

$$\mu_i = -\left.\frac{\partial E(\mathbf{E})}{\partial E_i}\right|_{\mathbf{E}=\mathbf{0}} + \sum_I Z_I R_{I,i} \tag{17}$$

$$\alpha_{ij} = -\left.\frac{\partial^2 E(\mathbf{E})}{\partial E_i \partial E_j}\right|_{\mathbf{E}=\mathbf{0}} \tag{18}$$

$$\beta_{ijk} = -\left.\frac{\partial^3 E(\mathbf{E})}{\partial E_i \partial E_j \partial E_k}\right|_{\mathbf{E}=\mathbf{0}} \tag{19}$$

$$\chi_{ij} = -\left.\frac{\partial^2 E(\mathbf{B})}{\partial B_i \partial B_j}\right|_{\mathbf{B}=\mathbf{0}} \tag{20}$$

where **E** and **B** are the external electric and magnetic fields, respectively.

## III. COMPUTATIONAL METHODS

We numerically investigated the antisymmetry rules in CCSs of charge-neutral and iso-electronic BN-doped benzene, naphthalene, anthracene, and pyrene derivatives ($C_{6-2m}B_mN_mH_6$ ($m$ = 0−3), $C_{10-2m}B_mN_mH_8$ ($m$ = 0−5), $C_{14-2m}B_mN_mH_{10}$ ($m$ = 0−7), and $C_{16-2m}B_mN_mH_{10}$ ($m$ = 0−8)). We used the Coulomb matrix[64] and bag-of-bonds[65] models implemented in the QML code[66] to obtain unique BN-doped PAH derivatives.[56] Alchemical enantiomers and diastereomers, connected *via* the reference pristine PAH, were identified from the Coulomb matrices for the BN dopants (Table SI). Geometries of all the derivatives are fixed to locally stable structures of pristine PAHs unless otherwise noted. The calculations of the singlet ground states were carried out.

Here, we describe the equations used as the antisymmetry rules for the considered BN-doped PAH derivatives. Eq. (9) was used for the electric dipole moments and hyperpolarizabilities of all derivatives. Eq. (10) was applied to the electric polarizabilities and magnetizabilities of BN-doped naphthalene, anthracene, and pyrene derivatives. For BN-doped benzene derivatives, the relation $P_{\mathrm{L},ij} \approx P_{\mathrm{R},ij}$ was employed for all the off-diagonal elements as well as the diagonal element along the rotational axis, while Eq. (12) corresponds to the other diagonal elements due to the rotational symmetry of reference benzene.

Kohn–Sham DFT (KSDFT) calculations were performed using the PBE0 functional.[67] Benchmark studies of DFT functionals indicate that PBE0 is accurate for electric dipole moments and polarizabilities.[68, 69] While other functionals may yield more accurate results for magnetizabilities, PBE0 performs reasonably well,[70] allowing us to use it consistently for all the properties. Our assessment of the accuracy of the antisymmetry rules was carried out within KSDFT using our chosen density functional approximation. The employed basis sets were pc-2[71, 72] for hydrogen and pcX-2[73] for other atoms, chosen to minimize the basis set error in the alchemical derivatives.[23, 24, 34, 74, 75] The gauge-including atomic orbitals (GIAOs) were used in the calculation of the magnetizabilities to address the gauge dependence issue. APDFT calculations were performed with the analytical alchemical derivatives within KSDFT.[23, 76] The derivation of the electric dipole moment and polarizabilities in APDFT is provided in the supplementary material. A modified version of an analytical APDFT code[75, 76] based on PySCF[77, 78] was used. Libcint[79] was used for computing atomic orbital integrals. The exchange–correlation functional was calculated with Libxc.[80, 81] The KSDFT calculations were conducted using Gaussian 16 rev. C02.[82] Figure 1(a) was created using POV-Ray.[83] The molecular and nanocluster structures were visualized with VMD.[84]

Further computational details are provided in the supplementary material.

## IV. RESULTS AND DISCUSSION

Firstly, we illustrate the antisymmetry rules of response properties (electric dipole moment, polarizabilities, hyperpolarizabilities, and magnetizabilities) with representative examples. Secondly, we statistically investigate the applicability and limitations of the antisymmetry rules for approximately 3,100 BN-doped PAH derivatives, including alchemical enantiomers and diastereomers. Subsequently, we demonstrate antisymmetry rule-based predictions of response properties. Finally, to elucidate the origin of the predictive power, we compare the numerical results of the antisymmetry rules and APDFT.

The antisymmetry rules for alchemical enantiomers of BN-doped naphthalene derivatives are illustrated in Fig. 1(c). We compared the relative magnitudes and orientations of the response properties predicted by the antisymmetry rules and KSDFT. The results show that, consistent with the antisymmetry rules, the magnitudes are approximately the same, and the orientations are nearly identical and opposite for even-order response properties (electric polarizabilities and magnetizabilities) and odd-order response properties (electric dipole moment and hyperpolarizabilities), respectively. We have investigated the relationships when the same alchemical enantiomers are in their locally stable structures (Fig. S1). The results show that geometry relaxation, which is not considered in our theory, leads to higher deviations from the antisymmetry rules, but the rules are still reasonably well satisfied. A realistic reference molecule is used in this case, while virtual materials with non-integer nuclear charges can also serve as reflection planes. Fig. 1(d) shows alchemical enantiomers of BN-doped naphthalene derivatives connected with a virtual reference compound. The electric dipole moments and polarizabilities are also close to the predictions of the antisymmetry rules. In both cases, the electronic energies exhibit approximate degeneracy, following alchemical chirality.

We statistically investigated distributions of the response properties of the relative amplitudes and angles among CCS (Figs. 2 and S2–S4). When pristine PAHs are reflection planes, most BN-doped PAH derivatives are classified as alchemical enantiomers or diastereomers (Table SI). We compared the medians of these distributions (Table SII). The alchemical enantiomers and diastereomers generally follow the antisymmetry rules well when compared to randomly selected pairs. For the electric polarizabilities and magnetizabilities, the alchemical enantiomers exhibit a closer agreement with the predictions of the antisymmetry rules than the alchemical diastereomers. This is because the antisymmetry rules for alchemical diastereomers are not exact in this case. In particular, the deviations in the magnetizabilities of the alchemical diastereomers are notably large, and a few outliers are found. Surprisingly, however, there is no significant difference in the electric polarizabilities.

Unlike the response properties, the electronic energy differences between pairs of the alchemical diastereomers are greater than those between randomly selected pairs (Fig. S5 and Table SIII), highlighting the importance of the careful selection of approximate alchemical enantiomers. A more general and accurate estimation requires the calculation of first-order alchemical derivatives.[52]

The response properties of one compound in an alchemical enantiomeric or diastereomeric pair can be approximately predicted from those of the other based on the antisymmetry rules. The computational cost of this approach is negligible. We assess the accuracy of the antisymmetry rule-based predictions by comparing them with the ground-truth KSDFT results (Figs. 3, S6, and S7). Overall, despite their simplicity, the predictive power is quite reasonable. In particular, the mean absolute errors (MAEs) for the electric dipole moments and polarizabilities are notably small. For the electric hyperpolarizabilities and magnetizabilities, a few pairs exhibit significantly large MAEs (Fig. S8). The accuracy of the magnetizabilities of the alchemical diastereomers is lower, but there are a few errors in the signs of the elements with large values (Fig. S6).

Further comparison between theoretical expectations and numerically observed trends (Figs. 3, S6, and S7) reveals additional insights. For electric dipole moments and hyperpolarizabilities, theory predicts no significant difference in the accuracy of the antisymmetry rules when applied to alchemical enantiomers versus alchemical diastereomers. One might assume that the energy difference/degeneracy in alchemical diastereomers/enantiomers would lead to larger errors of other property estimates for the former than for the latter. However, the accuracy in the alchemical predictions is rather related to how dramatic the changes in the electron density are (for energy, they are only reflected in the third- and higher-order terms of the perturbation expansion). Thus, a mere energy degeneracy in the case of alchemical enantiomers does not yet necessarily imply that the density changes are less dramatic. Indeed, in some cases, the MAEs for alchemical diastereomers are even lower than those for alchemical enantiomers. However, for the hyperpolarizabilities of BN-doped anthracene derivatives, two extreme outliers among the alchemical diastereomers increase the MAE (Fig. S8), deviating from the theoretically predicted trend. For electric polarizabilities and magnetizabilities, the computational results are consistent with theoretical expectations, showing that the MAEs for alchemical enantiomers are lower than those for alchemical diastereomers.

We have introduced Eqs. (9) and (10) as antisymmetry rules for the off-diagonal elements of second-order response properties. In the latter, we assume that the first-order perturbation term is relatively small. We compared the accuracy of Eqs. (9) and (10) (Table SIV). Eq. (10) shows better accuracy for the electric polarizabilities, while the magnetizabilities are almost unchanged. Therefore, Eq. (10) was adopted throughout this study.

The performance of ranking compounds based on the antisymmetry rules is evaluated using the Spearman correlation coefficients (SCCs) of magnitudes of the response properties (Table SV). The electric dipole moments and polarizabilities show remarkable correlations with SCCs greater than 0.94. The prediction accuracy of the antisymmetry rule for the electric hyperpolarizabilities is somewhat lower but still shows good performance (SCCs of ~0.90 for BN-doped naphthalene, anthracene, and pyrene derivatives). For the electric hyperpolarizabilities, there is no significant difference in SCCs between the alchemical enantiomers and diastereomers. SCCs for the magnetizabilities of the alchemical enantiomers and diastereomers are ~0.75 and ~0.45, respectively. These values are significantly lower than those for the other properties, and the difference between the alchemical enantiomers and diastereomers is substantial. The application of the antisymmetry rules to ranking the magnetizabilities of alchemical diastereomers may be challenging.

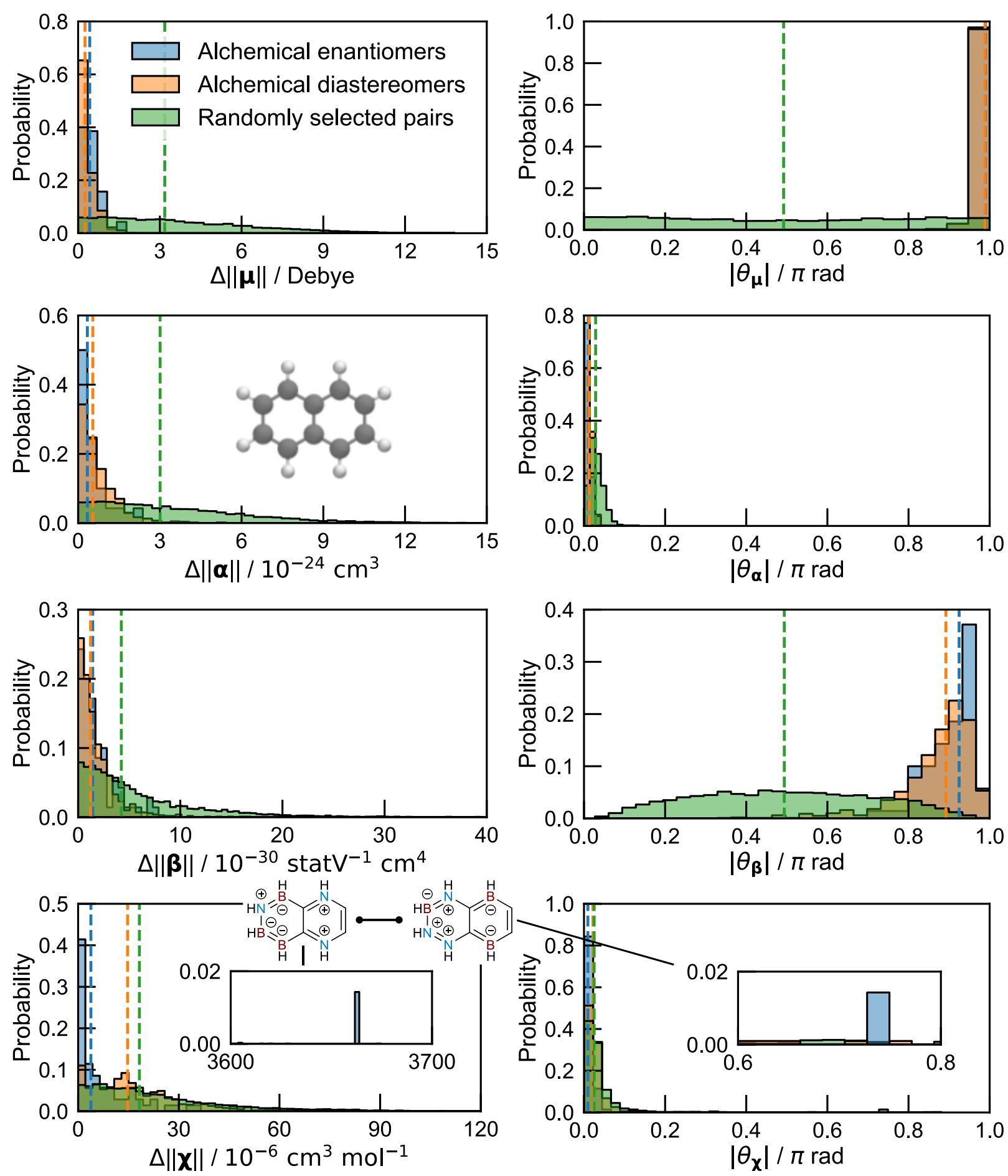

**FIG 2.** Probability distributions of the relative electric dipole moments (**μ**), polarizabilities (**α**), hyperpolarizabilities (**β**), and magnetizabilities (**χ**) between alchemical enantiomers, alchemical diastereomers, and 10,000 randomly selected pairs for all 2,286 BN-doped naphthalene derivatives, including one reference. $\Delta\|\mathbf{P}\|$ is the absolute relative magnitude of property **P**. $|\theta_{\mathbf{P}}|$ is the absolute relative angle (in $\pi$ radians) between the response properties. The broken lines represent the medians. For visibility, a few outliers of the alchemical diastereomers and randomly selected pairs are outside the range.

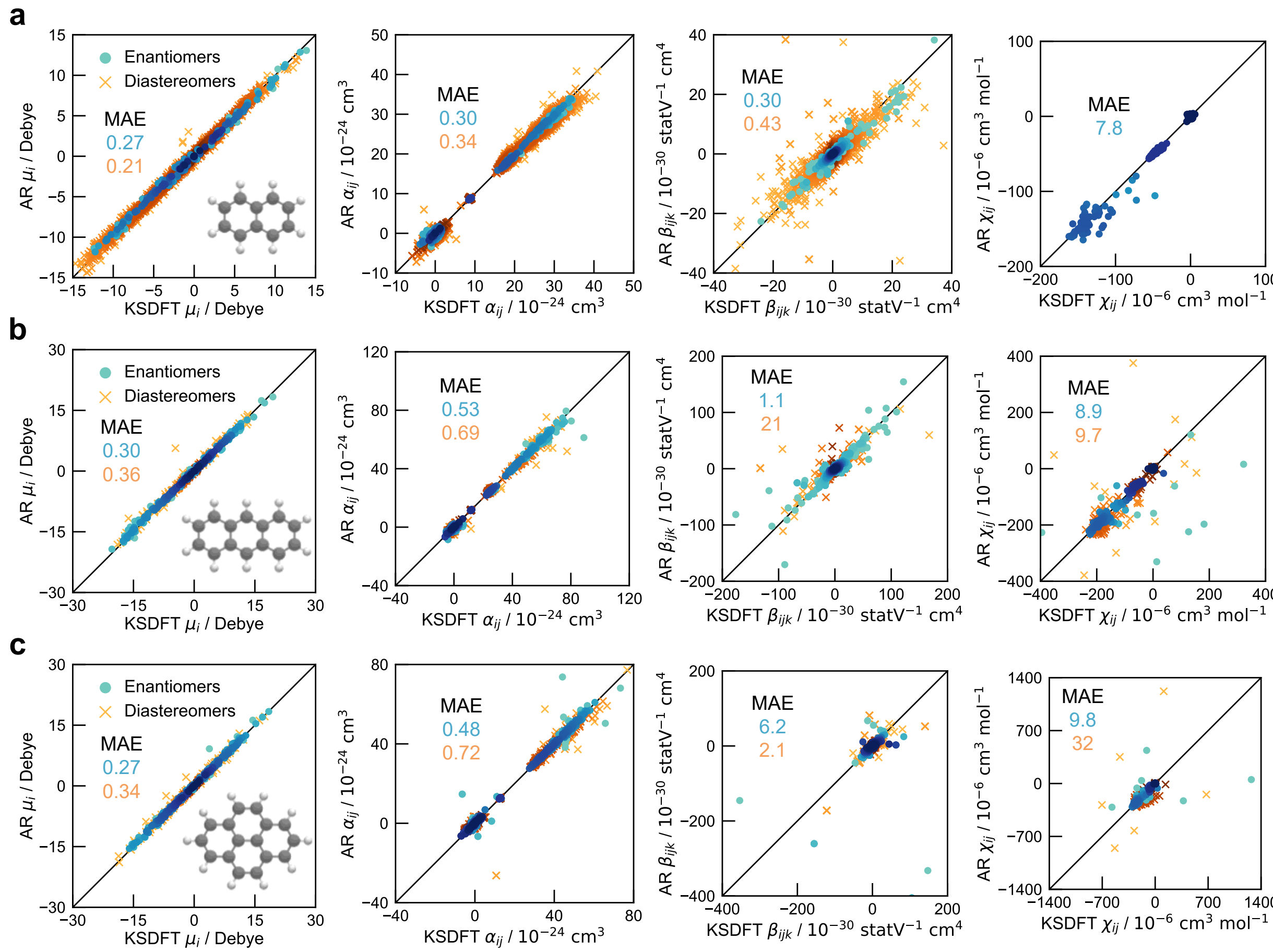

**FIG 3.** Antisymmetry rules (ARs)-based predictions of the electric dipole moments (**μ**), polarizabilities (**α**), hyperpolarizabilities (**β**), and magnetizabilities (**χ**) of the alchemical enantiomers and diastereomers for BN-doped (a) naphthalene, (b) anthracene, and (c) pyrene derivatives, compared with the results of KSDFT. The intensity of the color represents the base-10 logarithm density. *i*, *j*, and *k* are one of the axes of the Cartesian coordinates. The results for the magnetizabilities of the alchemical diastereomers of the BN-doped naphthalene derivatives are shown in Fig. S6. For visibility, a few outliers for the electric hyperpolarizabilities and magnetizabilities are outside the range. All the data are shown in Fig. S8.

We numerically evaluate the accuracy of APDFT and investigate the origin of the predictive power of the antisymmetry rules. The antisymmetry rules have been derived from APDFT. These rules are obtained within the first- and second-order perturbation expansions (APDFT1 and APDFT2) for odd-order response properties (electric dipole moment and hyperpolarizabilities) and even-order response properties (electric polarizabilities and magnetizabilities), respectively. Unlike the antisymmetry rules, the scope of APDFT encompasses not only antisymmetric pairs but also all compounds adjacent to a reference material in CCS. In addition, the symmetry requirement for the reference is lifted in APDFT. Here, we compare the antisymmetry rules with APDFT for the electric dipole moments and polarizabilities. APDFT1 and KSDFT dipole moments are in good agreement (Figs. 4(a) and S9), indicating the importance of the first-order perturbation term. MAEs of the antisymmetry rule-based predictions (Figs. 3 and S7) are much smaller than those of APDFT1. This difference could be attributed to the ignored higher-order terms, assuming the convergence of the perturbation expansion. The accuracy of APDFT1 polarizabilities is relatively low (Fig. 4(a) and S9), and the corresponding antisymmetry rule derived based on APDFT2 outperforms APDFT1. Next, we investigate the relative response properties of the alchemical enantiomers and diastereomers at the APDFT2 level from Eq. (6) without implementing higher-order energy derivatives (Figs. 4(b) and S10). The overall prediction trends for the dipole moments are unchanged. On the other hand, MAEs for the relative polarizabilities decrease significantly, showing the large contribution of the even-order perturbation terms. Thus, the antisymmetry rules properly incorporate the important contributions in APDFT.

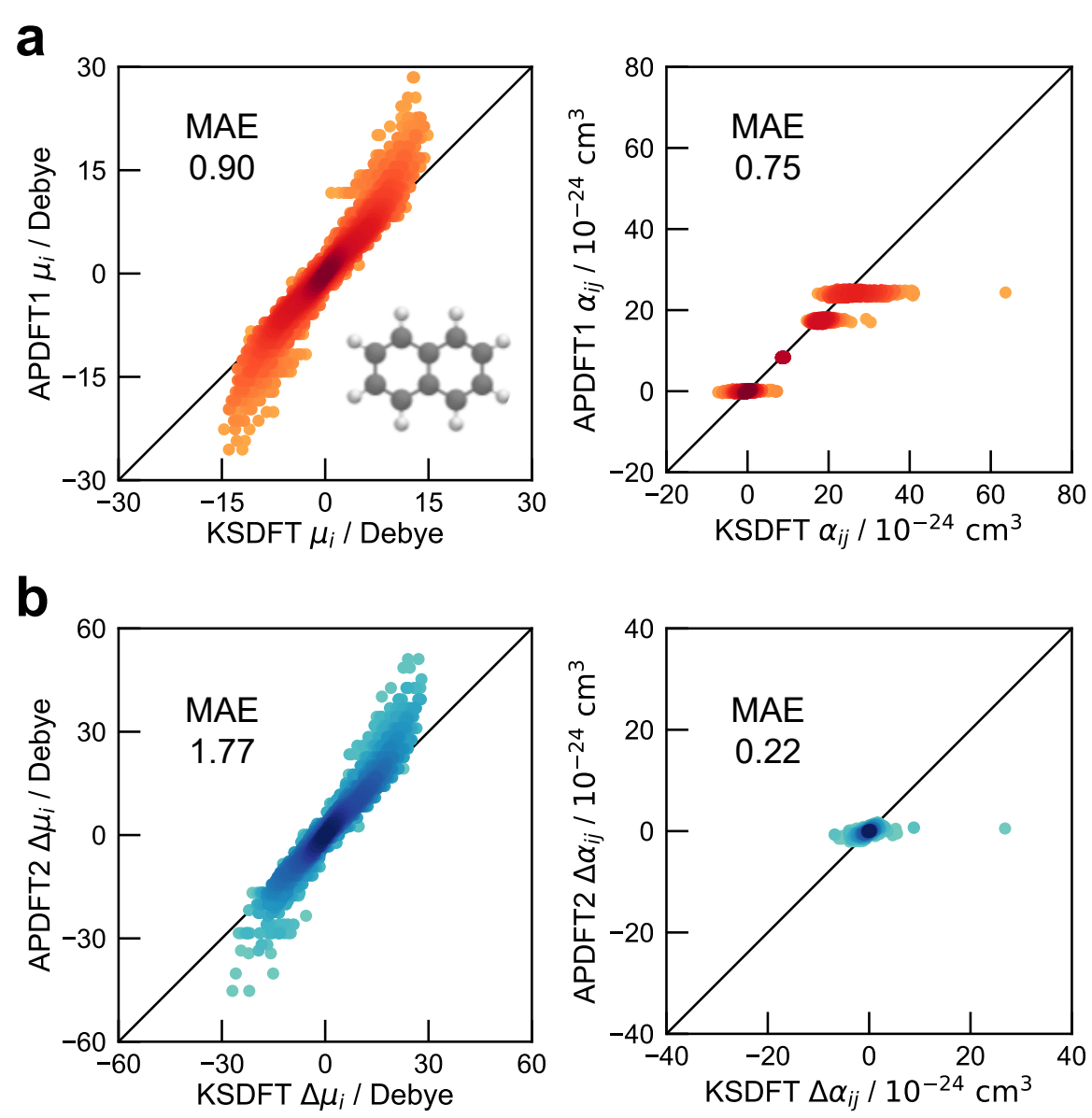

**FIG 4.** Comparisons between APDFT and KSDFT predictions of the electric dipole moments (**μ**) and polarizabilities (**α**) for BN-doped naphthalene derivatives. (a) APDFT1 predictions of the dipole moments and polarizabilities for all 2286 derivatives, including one reference. (b) APDFT2 predictions of the relative dipole moments and polarizabilities of the alchemical enantiomers and diastereomers. The intensity of the color represents the base-10 logarithm density. $i$ and $j$ are one of the axes of the Cartesian coordinates.

## V. CONCLUSIONS

In this work, we introduce antisymmetry rules for the response properties of alchemical enantiomers and diastereomers under uniform static external fields. These rules have been obtained from quantum alchemy and symmetry considerations and are universal for various CCSs within the perturbation approximations used. We numerically investigated the applicability and limitations for electric dipole moments, polarizabilities, hyperpolarizabilities, and magnetizabilities of approximately 3,100 charge-neutral and iso-electronic BN-doped PAH derivatives. Our statistical analysis shows that despite some discrepancies in magnetizabilities, the relative response properties of most alchemical enantiomers and diastereomers are amenable to the antisymmetry rules. The antisymmetry rules provide a simple and efficient way to estimate the response properties of one of two materials in an alchemical enantiomeric or diastereomeric pair from those of the other. We show that the antisymmetry rule-based predictions are reasonably accurate compared with those of the ground-truth quantum chemical method. The operations required to make these predictions are intuitive, and no additional experiments or quantum chemistry calculations are needed. Furthermore, the APDFT electric dipole moments and polarizabilities are compared with the antisymmetry rule-based predictions, highlighting the importance of the leading order terms in the perturbation expansion, which are properly incorporated in the antisymmetry rules. The antisymmetry rules can be used to deepen our understanding of CCS and to narrow down its important domain.

## DATA AVAILABILITY

The code used for upstream processing and reproducing the data is available on Zenodo (https://doi.org/10.5281/zenodo.14882369), along with all necessary intermediate data. The code used for quantum alchemy calculations and analyses is available on Zenodo (https://doi.org/10.5281/zenodo.14885475) and GitHub (https://github.com/takafumi-shiraogawa/AQA).

## SUPPLEMENTARY MATERIAL

Computational details; supplementary data (Figs. S1–S10, Tables SI–SV, and the locally stable geometries of the pristine PAHs); spatial symmetry of electron density derivatives with respect to static and uniform external fields; derivation of electric dipole moments and polarizabilities in APDFT.

## ACKNOWLEDGMENTS

T.S. acknowledges the support from the Japan Society for the Promotion of Science (JSPS) grant No. JP21J00210. M.E. acknowledges the support from JSPS grant-in-aid for Transformative Research Areas (A) (JP22H05133 and JP22H05131). O.A.v.L. received the funding from the European Research Council (ERC) under the European Union's Horizon 2020 Research and Innovation Program (Grant Agreement No. 772834). This research was part of the University of Toronto's Acceleration Consortium, which received the funding from the Canada First Research Excellence Fund (CFREF). O.A.v.L. received the support as the Ed Clark Chair of Advanced Materials and as a Canada CIFAR AI Chair. O.A.v.L. also acknowledges the support of the Natural Sciences and Engineering Research Council of Canada (NSERC), RGPIN-2023-04853. The computations were partly performed at the Research Center for Computational Science, Okazaki, Japan (23-IMS-C197 and 24-IMS-C194).

## AUTHOR DECLARATIONS

### Conflict of Interest

The authors have no conflicts to disclose.

### Author Contributions

**Takafumi Shiraogawa:** Conceptualization (lead); Data curation (lead); Methodology (lead); Software (lead); Investigation (lead); Writing– original draft (lead); Writing– review & editing (equal). **Simon León Krug:** Methodology (supporting); Investigation (supporting); Writing– review & editing (equal). **Masahiro Ehara:** Conceptualization (supporting); Methodology (supporting); Investigation (supporting); Supervision (equal); Writing– review & editing (equal). **O. Anatole von Lilienfeld:** Conceptualization (supporting); Methodology (supporting); Investigation (supporting); Supervision (equal); Writing– review & editing (equal).

# Supplementary Material

# Antisymmetry rules of response properties in certain chemical spaces

Takafumi Shiraogawa,[1,2,3,a)] Simon León Krug,[4] Masahiro Ehara,[1,2,3,b)]
and O. Anatole von Lilienfeld[4,5,6,7,8,9,10,c)]

*[1]Institute for Molecular Science, National Institutes of Natural Sciences, 38 Nishigonaka, Myodaiji, Okazaki 444-8585, Japan*

*[2]Research Center for Computational Science, National Institutes of Natural Sciences, 38 Nishigonaka, Myodaiji, Okazaki 444-8585, Japan*

*[3]The Graduate University for Advanced Studies, 38 Nishigonaka, Myodaiji, Okazaki 444-8585, Japan*

*[4]Machine Learning Group, Technische Universität Berlin, 10587 Berlin, Germany*

*[5]Chemical Physics Theory Group, Department of Chemistry, University of Toronto, St. George Campus, Toronto, M5S3H6 Ontario, Canada*

*[6]Department of Materials Science and Engineering, University of Toronto, St. George Campus, Toronto, M5S 3E4 Ontario, Canada*

*[7]Vector Institute for Artificial Intelligence, Toronto, M5S 1M1 Ontario, Canada*

*[8]Berlin Institute for the Foundations of Learning and Data, 10587 Berlin, Germany*

*[9]Department of Physics, University of Toronto, St. George Campus, Toronto, M5S 1A7 Ontario, Canada*

*[10]Acceleration Consortium, University of Toronto, Toronto, M5R 0A3 Ontario, Canada*

Correspondence to: a) shiraogawa@ims.ac.jp; b) ehara@ims.ac.jp; c) anatole.vonlilienfeld@utoronto.ca

## Contents

- Computational details
- Supplementary data
- Spatial symmetry of electron density derivatives with respect to static and uniform external fields
- Derivation of electric dipole moment and polarizabilities in APDFT
- References in the supplementary material

## Computational details

To reduce the computational cost, for the BN-doped anthracene and pyrene derivatives, 100 pairs each of alchemical enantiomers and alchemical diastereomers were randomly chosen and calculated. 10,000 randomly selected pairs from these do not contain alchemical enantiomers or diastereomers. The angles between tensors were calculated based on the Frobenius product. To define the angles, any pair with a vector or tensor where all elements are zero was excluded from the relative angle estimations. For the polarizabilities and magnetizabilities of BN-doped benzene derivatives, diagonal elements $P_{\mathrm{L},ii}^{(2)}$ and $P_{\mathrm{L},jj}^{(2)}$ are swapped so that the angle becomes zero when the antisymmetry rule is fulfilled. The order or relative alignment of the randomly selected molecules is not guaranteed. The electric dipole moments, polarizabilities, and hyperpolarizabilities were calculated at a geometric center of reference molecules.

The locally stable structures of the pristine PAHs are provided in the supplementary data section.

## Supplementary data

TABLE SI. The number of BN-doped PAH derivatives and alchemical enantiomers and diastereomers. Pristine PAHs are included.

| Reference PAH | BN-doped PAH derivatives | Pairs of alchemical enantiomers | Pairs of alchemical diastereomers |
|---|---|---|---|
| benzene | 18 | 3 | 0 |
| naphthalene | 2286 | 70 | 997 |
| anthracene | 154422 | 1951 | 74539 |
| pyrene | 1301055 | 6022 | 642463 |

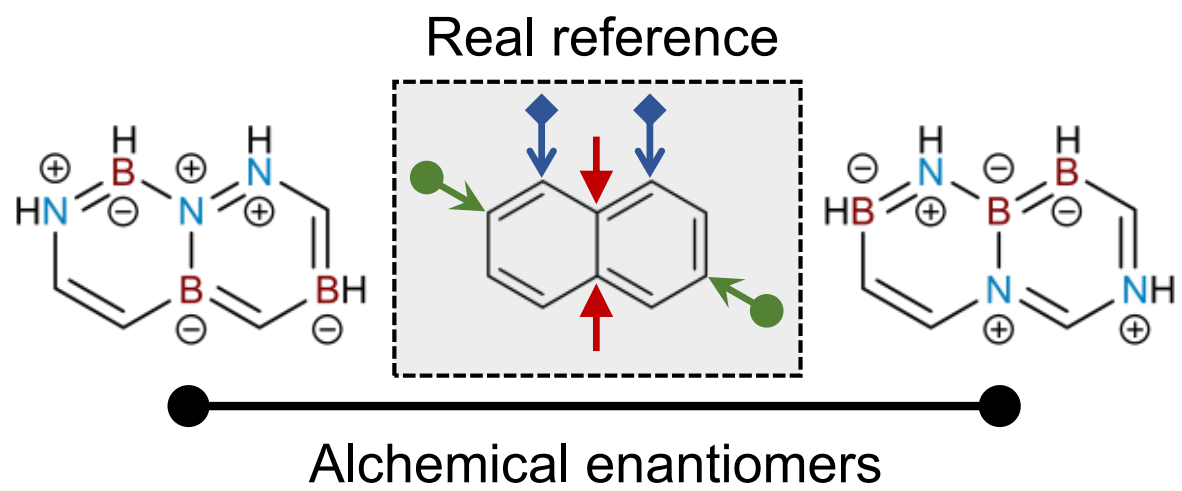

$\widetilde{|\Delta E|}$ = 0.173 (AR: 0.000)

$\overline{\Delta\|\boldsymbol{\mu}\|}$ = 0.11, $|\theta_{\boldsymbol{\mu}}|$ = 0.92 (AR: 0.00, 1.00)

$\overline{\Delta\|\boldsymbol{\alpha}\|}$ = 0.02, $|\theta_{\boldsymbol{\alpha}}|$ = 0.02 (AR: 0.00, 0.00)

$\overline{\Delta\|\boldsymbol{\beta}\|}$ = 0.24, $|\theta_{\boldsymbol{\beta}}|$ = 0.78 (AR: 0.00, 1.00)

$\overline{\Delta\|\boldsymbol{\chi}\|}$ = 0.02, $|\theta_{\boldsymbol{\chi}}|$ = 0.00 (AR: 0.00, 0.00)

FIG. S1 Antisymmetry rules (ARs) for a pair of alchemical enantiomers of BN-doped naphthalene derivatives. The geometries were relaxed to the locally stable structures. The geometries are maximally overlapped by translation and rotation. The absolute relative magnitude of the property $\mathbf{P}$, $\widetilde{\Delta\|\mathbf{P}\|}$, is divided by the corresponding median for randomly selected pairs. $\overline{\Delta\|\mathbf{P}\|}$ denotes the absolute relative property divided by the average of the respective absolute values. $|\theta_{\mathbf{P}}|$ is the absolute relative angle (in $\pi$ radians) between the response properties.

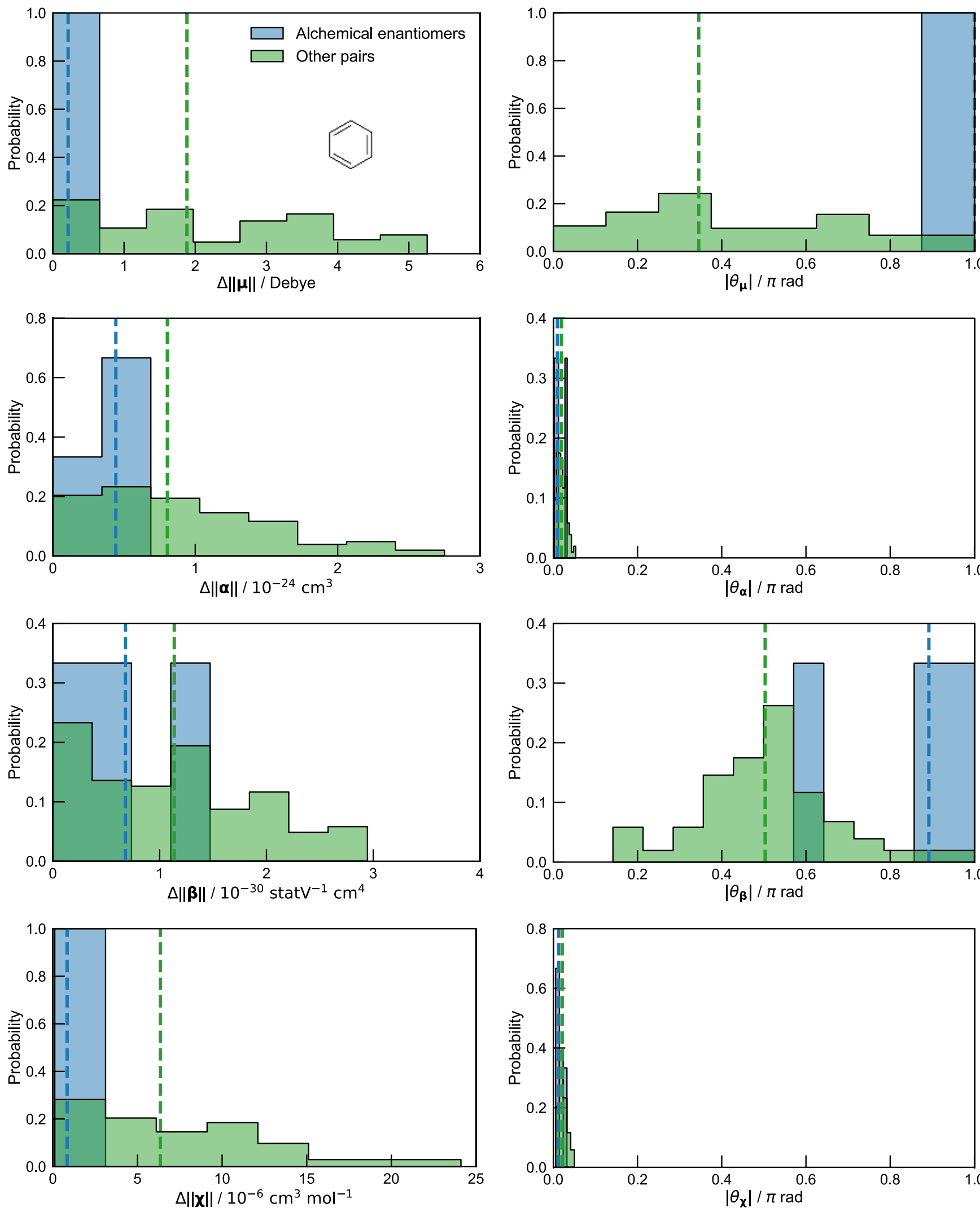

FIG. S2. Probability distributions of the relative electric dipole moments ($\boldsymbol{\mu}$), polarizabilities ($\boldsymbol{\alpha}$), hyperpolarizabilities ($\boldsymbol{\beta}$), and magnetizabilities ($\boldsymbol{\chi}$) between alchemical enantiomers and all other pairs of BN-doped benzene derivatives. $\Delta\|\mathbf{P}\|$ is the absolute relative magnitude of property $\mathbf{P}$. $|\theta_{\mathbf{P}}|$ is the absolute relative angle (in $\pi$ radians) between the response properties. The broken lines represent the medians.

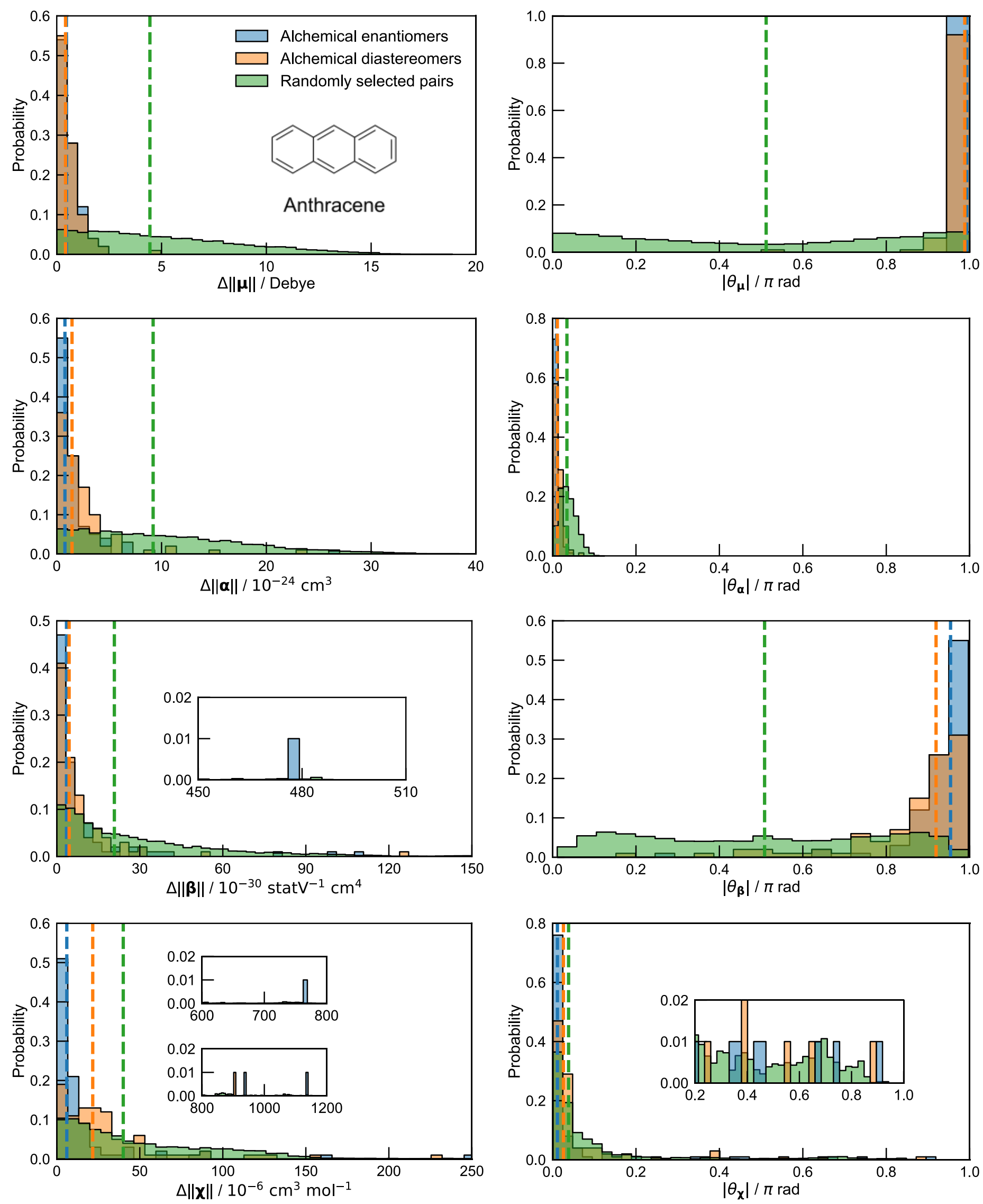

FIG. S3. Probability distributions of the relative electric dipole moments (**μ**), polarizabilities (**α**), hyperpolarizabilities (**β**), and magnetizabilities (**χ**) between alchemical enantiomers, alchemical diastereomers, and 10,000 randomly selected pairs of BN-doped anthracene derivatives. $\Delta||\mathbf{P}||$ is the absolute relative magnitude of property **P**. $|\theta_{\mathbf{P}}|$ is the absolute relative angle (in $\pi$ radians) between the response properties. The broken lines represent the medians. For visibility, a few outliers of the alchemical diastereomers and randomly selected pairs are outside the range.

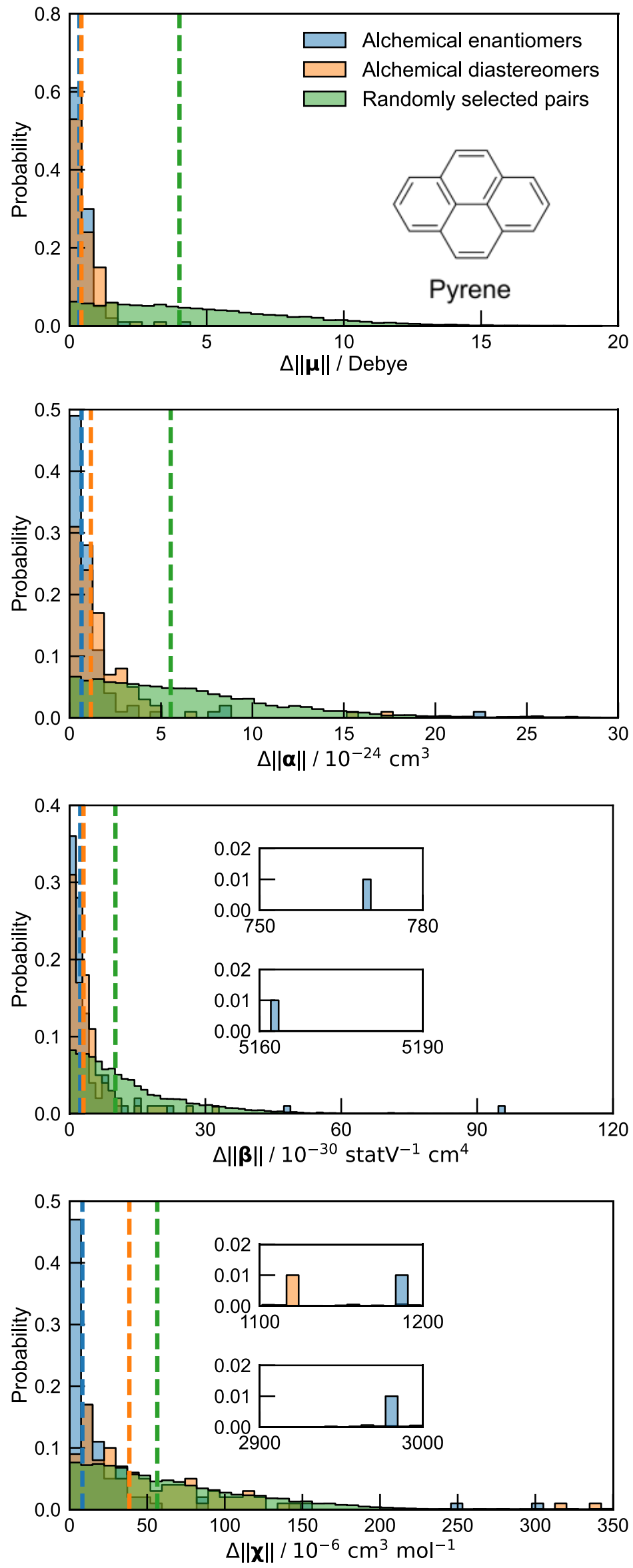

FIG. S4. Probability distributions of the relative electric dipole moments (**μ**), polarizabilities (**α**), hyperpolarizabilities (**β**), and magnetizabilities (**χ**) between alchemical enantiomers, alchemical diastereomers, and 10,000 randomly selected pairs of BN-doped pyrene derivatives. $\Delta||\mathbf{P}||$ is the absolute relative magnitude of property **P**. $|\theta_{\mathbf{P}}|$ is the absolute relative angle (in $\pi$ radians) between the response properties. The broken lines represent the medians. For visibility, a few outliers of the alchemical diastereomers and randomly selected pairs are outside the range.

TABLE SII. Medians of probability distributions of the relative electric dipole moments (**μ**), polarizabilities (**α**), hyperpolarizabilities (**β**), and magnetizabilities (**χ**) between alchemical enantiomers, alchemical diastereomers, and other pairs for BN-doped PAH derivatives (please see Figs. 2 and S2−S4). For BN-doped naphthalene, anthracene, and pyrene derivatives, the results of 10,000 randomly selected pairs are compared with those of the alchemical enantiomers and diastereomers. The units of **μ**, **α**, **β**, and **χ** are Debye, $10^{-24}$ cm$^3$, $10^{-30}$ statV$^{-1}$ cm$^4$, and $10^{-6}$ cm$^3$ mol$^{-1}$, respectively.

| CCS | Property | $\Delta\|\mathbf{P}\|$ | | | $\|\theta_{\mathbf{P}}\|$ ($\pi$ rad) | | |
|---|---|---|---|---|---|---|---|
| | | Alchemical enantiomers | Alchemical diastereomers | Other pairs | Alchemical enantiomers | Alchemical diastereomers | Other pairs |
| BN-doped benzene derivatives | **μ** | 0.21 | - | 1.88 | 1.00 | - | 0.35 |
| | **α** | 0.44 | - | 0.80 | 0.01 | - | 0.02 |
| | **β** | 0.68 | - | 1.14 | 0.89 | - | 0.50 |
| | **χ** | 0.84 | - | 6.35 | 0.01 | - | 0.02 |
| BN-doped naphthalene derivatives | **μ** | 0.43 | 0.25 | 3.18 | 0.99 | 0.99 | 0.49 |
| | **α** | 0.34 | 0.54 | 3.01 | 0.01 | 0.01 | 0.03 |
| | **β** | 1.43 | 1.22 | 4.23 | 0.92 | 0.89 | 0.49 |
| | **χ** | 3.80 | 14.75 | 18.26 | 0.01 | 0.02 | 0.03 |
| BN-doped anthracene derivatives | **μ** | 0.46 | 0.42 | 4.45 | 0.99 | 0.99 | 0.51 |
| | **α** | 0.78 | 1.46 | 9.20 | 0.01 | 0.01 | 0.03 |
| | **β** | 3.52 | 4.47 | 20.81 | 0.95 | 0.92 | 0.51 |
| | **χ** | 6.05 | 21.67 | 40.05 | 0.01 | 0.03 | 0.04 |
| BN-doped pyrene derivatives | **μ** | 0.36 | 0.43 | 4.01 | 0.99 | 0.99 | 0.50 |
| | **α** | 0.66 | 1.17 | 5.53 | 0.01 | 0.01 | 0.03 |
| | **β** | 2.38 | 3.10 | 10.14 | 0.93 | 0.91 | 0.49 |
| | **χ** | 8.30 | 38.57 | 56.56 | 0.01 | 0.03 | 0.04 |

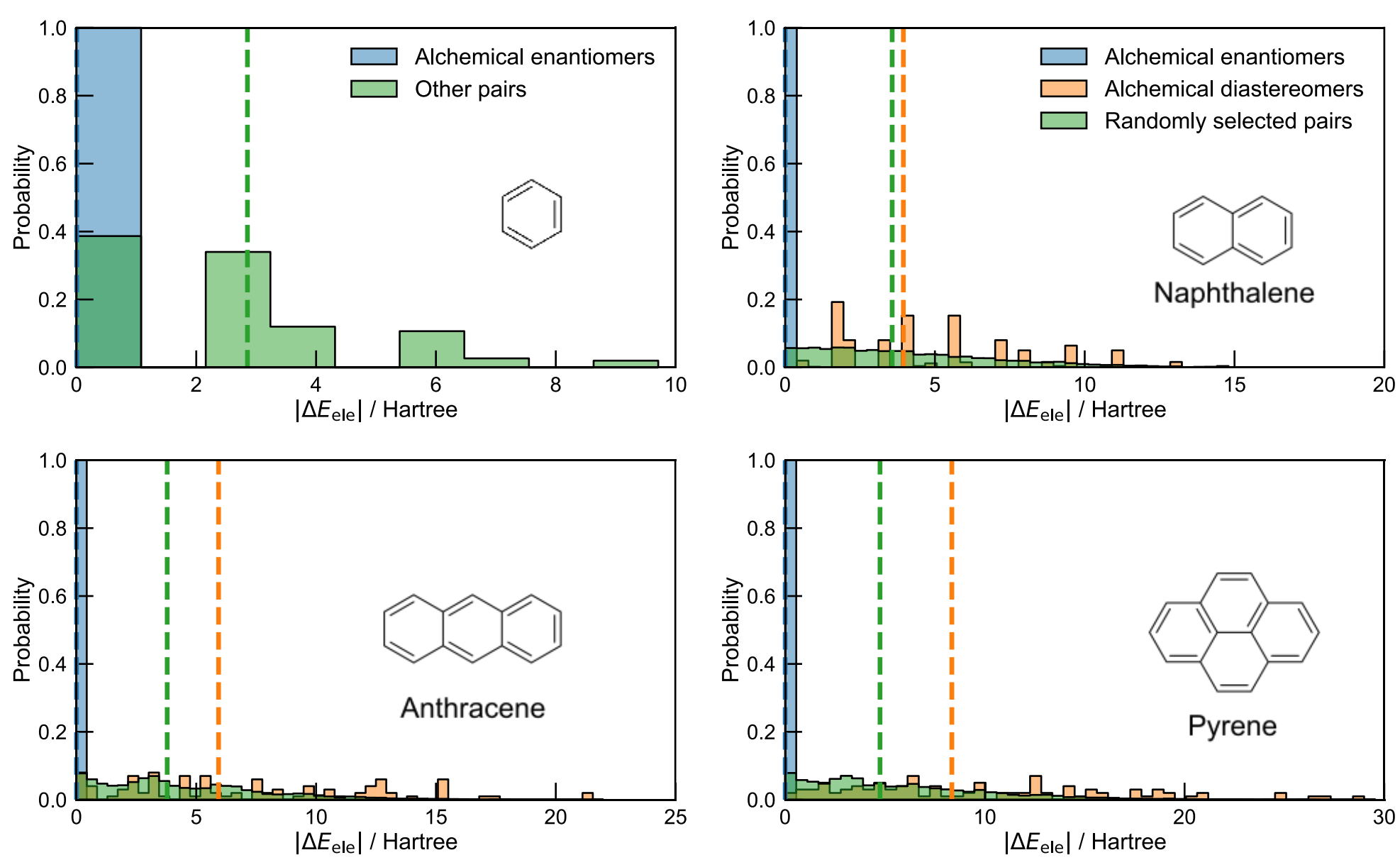

FIG. S5. Probability distributions of the relative electronic energies between alchemical enantiomers, alchemical diastereomers, and other pairs of BN-doped PAH derivatives. For BN-doped naphthalene, anthracene, and pyrene derivatives, the results of 10,000 randomly selected pairs are compared with those of the alchemical enantiomers and diastereomers. The broken lines represent the medians. For visibility, a few outliers of the alchemical diastereomers and randomly selected pairs are outside the range.

TABLE SIII. Medians of probability distributions of the relative electronic energies (in Hartree) between alchemical enantiomers, alchemical diastereomers, and other pairs of BN-doped PAH derivatives.

| CCS | Alchemical enantiomers | Alchemical diastereomers | Other pairs |
|---|---|---|---|
| BN-doped benzene derivatives | 0.006 | - | 2.854 |
| BN-doped naphthalene derivatives | 0.004 | 3.947 | 3.574 |
| BN-doped anthracene derivatives | 0.004 | 5.932 | 3.788 |
| BN-doped pyrene derivatives | 0.004 | 8.348 | 4.751 |

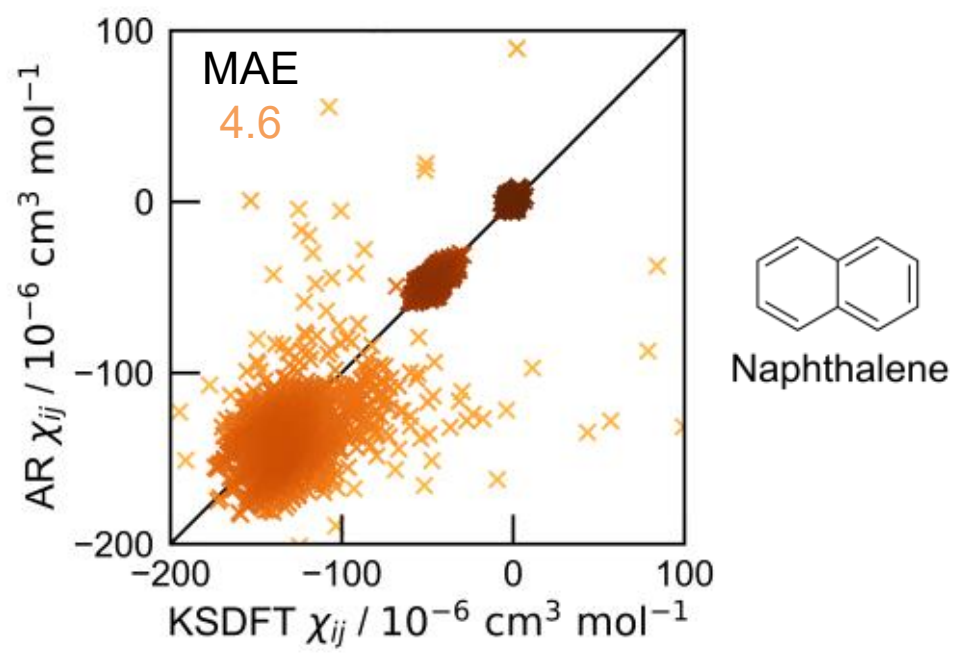

FIG. S6. Antisymmetry rule (AR)-based predictions of the magnetizabilities (**χ**) of alchemical diastereomers for all 2286 BN-doped naphthalene derivatives including one reference, compared with the results of KSDFT. This figure corresponds to Fig. 3. The intensity of the color represents the base-10 logarithm density. $i$, $j$, and $k$ are one of the axes of the Cartesian coordinates. For visibility, a few outliers are outside the range.

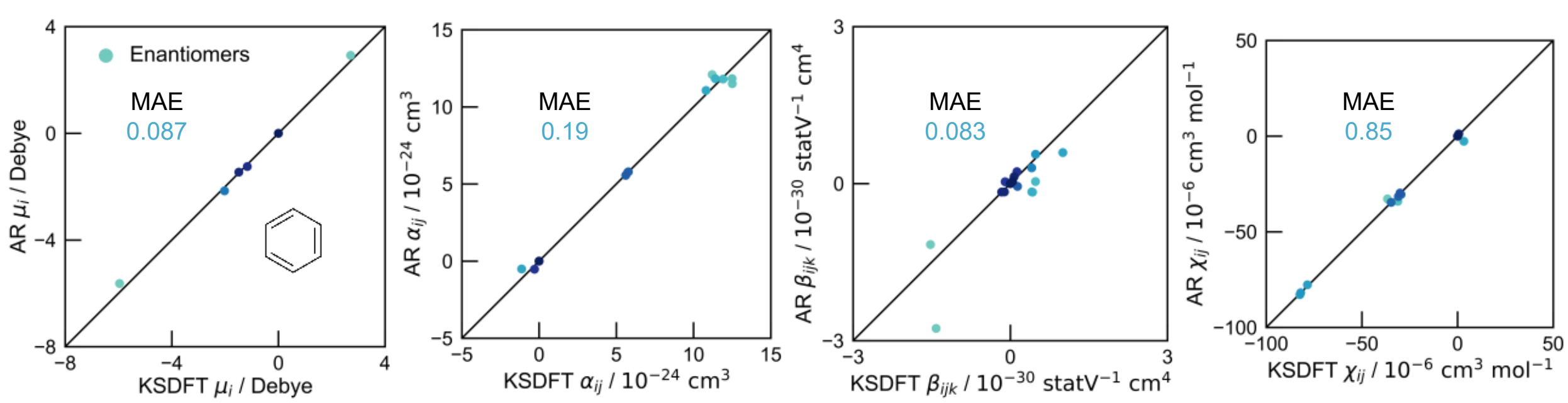

FIG. S7. Antisymmetry rules (ARs)-based predictions of the electric dipole moments (**μ**), polarizabilities (**α**), hyperpolarizabilities (**β**), and magnetizabilities (**χ**) of alchemical enantiomers for BN-doped benzene derivatives compared with the results of KSDFT. The intensity of the color represents the base-10 logarithm density. $i$, $j$, and $k$ are one of the axes of the Cartesian coordinates.

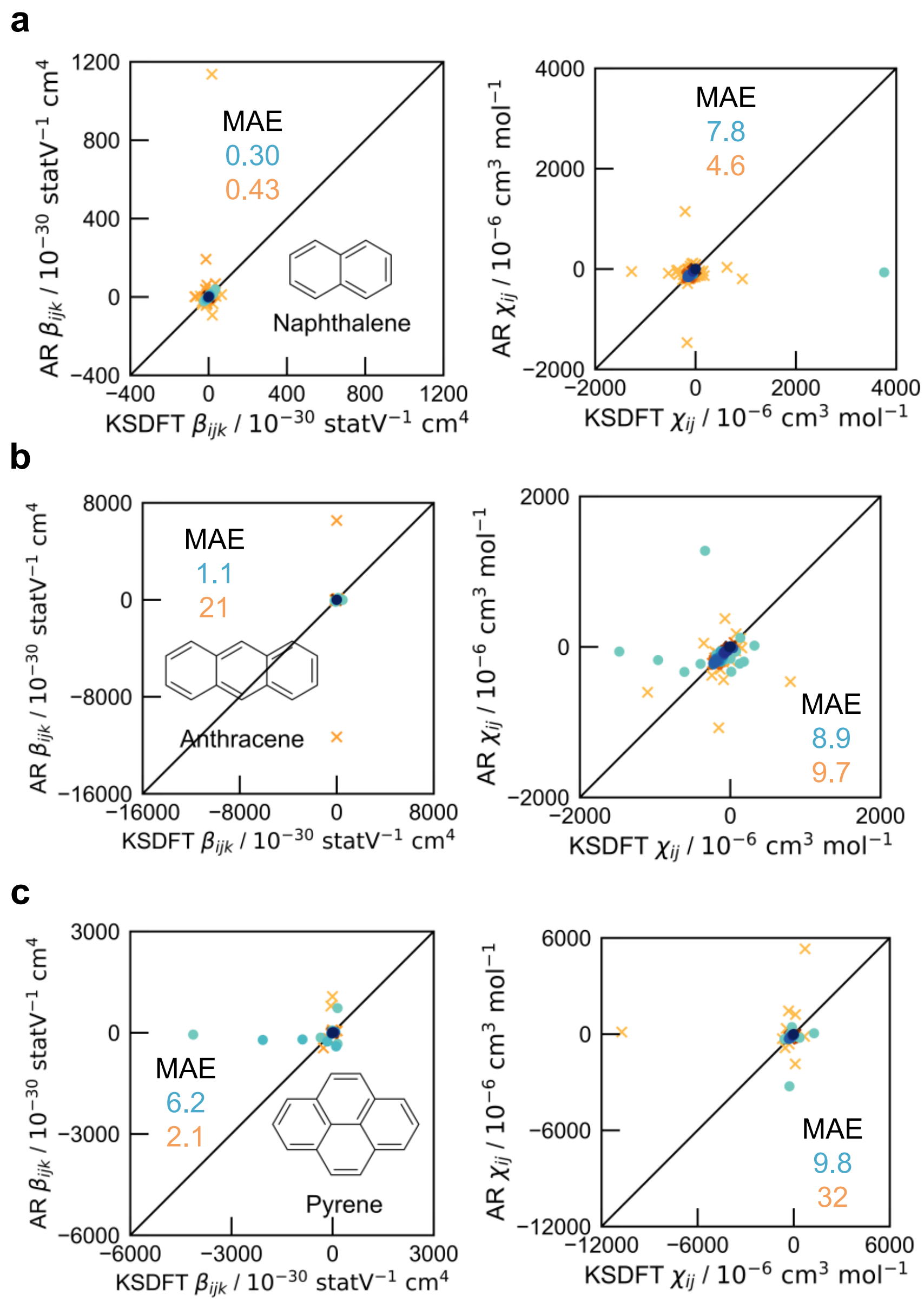

FIG. S8. Antisymmetry rules (ARs)-based predictions of the electric hyperpolarizabilities (**β**) and magnetizabilities (**χ**) of alchemical enantiomers and diastereomers for BN-doped (a) benzene, (b) anthracene, and (c) pyrene derivatives compared with the results of KSDFT. The intensity of the color represents the base-10 logarithm density. *i*, *j*, and *k* are one of the axes of the Cartesian coordinates.

TABLE SIV. MAEs of antisymmetry rules-based predictions of the electric polarizabilities (**α**) and magnetizabilities (**χ**) of alchemical enantiomers and diastereomers for BN-doped PAH derivatives compared with the results of KSDFT. The off-diagonal elements are evaluated based on different Eqs. (9) and (10) in the main text.

| CCS | Property | Eq. (9) | | Eq. (10) | |
|---|---|---|---|---|---|
| | | Alchemical enantiomers | Alchemical diastereomers | Alchemical enantiomers | Alchemical diastereomers |
| BN-doped benzene derivatives | **α** | 0.31 | - | 0.19 | - |
| | **χ** | 0.55 | - | 0.85 | - |
| BN-doped naphthalene derivatives | **α** | 0.49 | 0.59 | 0.30 | 0.34 |
| | **χ** | 7.83 | 4.61 | 7.85 | 4.56 |
| BN-doped anthracene derivatives | **α** | 0.93 | 1.14 | 0.53 | 0.69 |
| | **χ** | 8.96 | 9.85 | 8.92 | 9.74 |
| BN-doped pyrene derivatives | **α** | 1.09 | 1.19 | 0.48 | 0.72 |
| | **χ** | 9.78 | 32.43 | 9.80 | 32.49 |

TABLE SV. Spearman correlation coefficients of the antisymmetry rules-based predictions and KSDFT estimations of magnitudes of the electric dipole moments (**μ**), polarizabilities (**α**), hyperpolarizabilities (**β**), and magnetizabilities (||**μ**||, ||**α**||, ||**β**||, and ||**χ**||) for BN-doped PAH derivatives.

| CCS | Property | Alchemical enantiomers | Alchemical diastereomers |
|---|---|---|---|
| BN-doped benzene derivatives | **μ** | 1.00 | - |
| | **α** | 1.00 | - |
| | **β** | 0.50 | - |
| | **χ** | 1.00 | - |
| BN-doped naphthalene derivatives | **μ** | 0.99 | 0.99 |
| | **α** | 0.96 | 0.97 |
| | **β** | 0.93 | 0.89 |
| | **χ** | 0.68 | 0.40 |
| BN-doped anthracene derivatives | **μ** | 0.98 | 0.98 |
| | **α** | 0.97 | 0.94 |
| | **β** | 0.90 | 0.86 |
| | **χ** | 0.82 | 0.52 |
| BN-doped pyrene derivatives | **μ** | 0.99 | 0.97 |
| | **α** | 0.95 | 0.95 |
| | **β** | 0.90 | 0.91 |
| | **χ** | 0.74 | 0.44 |

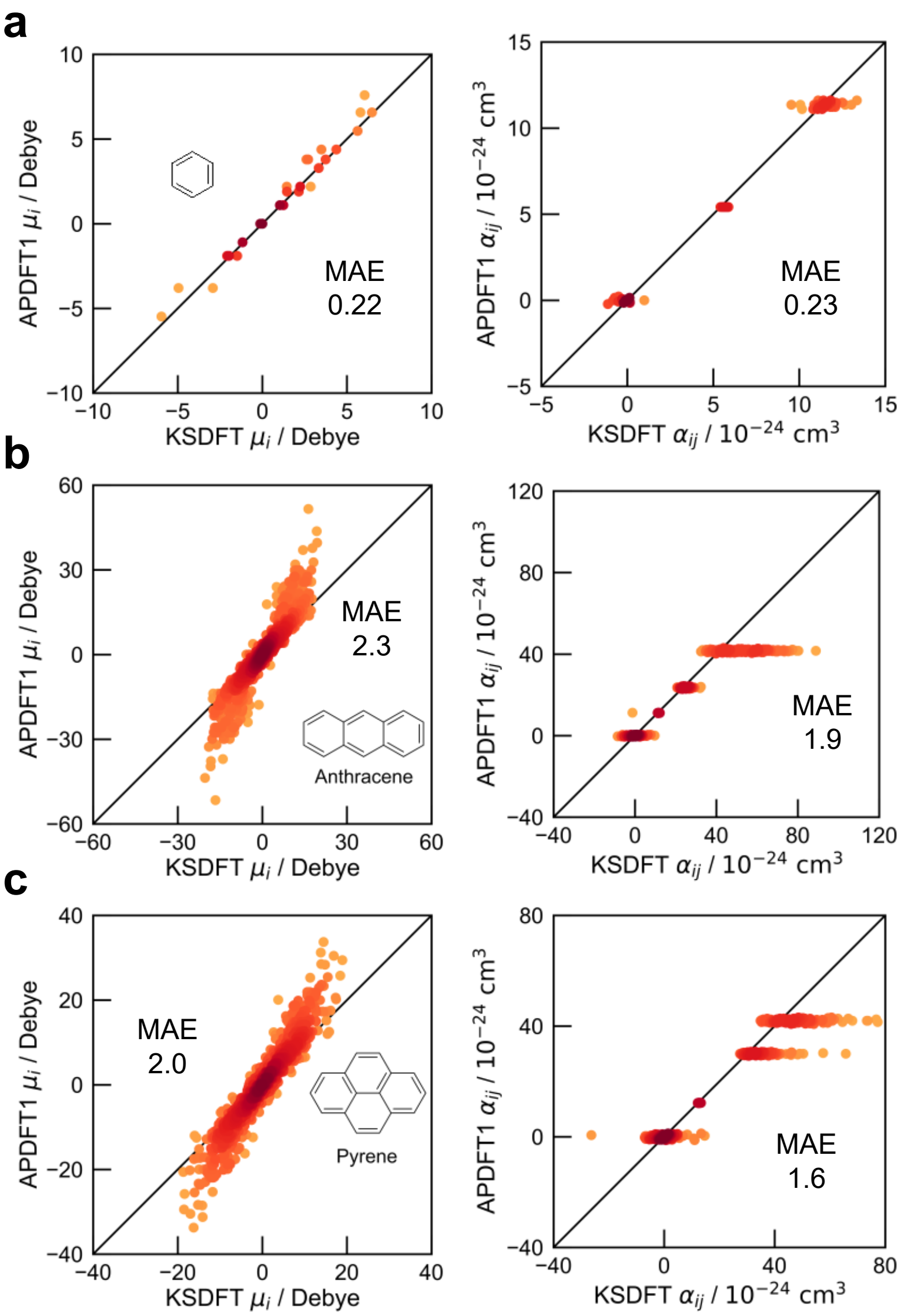

FIG. S9. APDFT1 predictions of the electric dipole moments (**μ**) and polarizabilities (**α**) of BN-doped (a) benzene, (b) anthracene, and (c) pyrene derivatives compared with the results of KSDFT. The intensity of the color represents the base-10 logarithm density. *i* and *j* are one of the axes of the Cartesian coordinates.

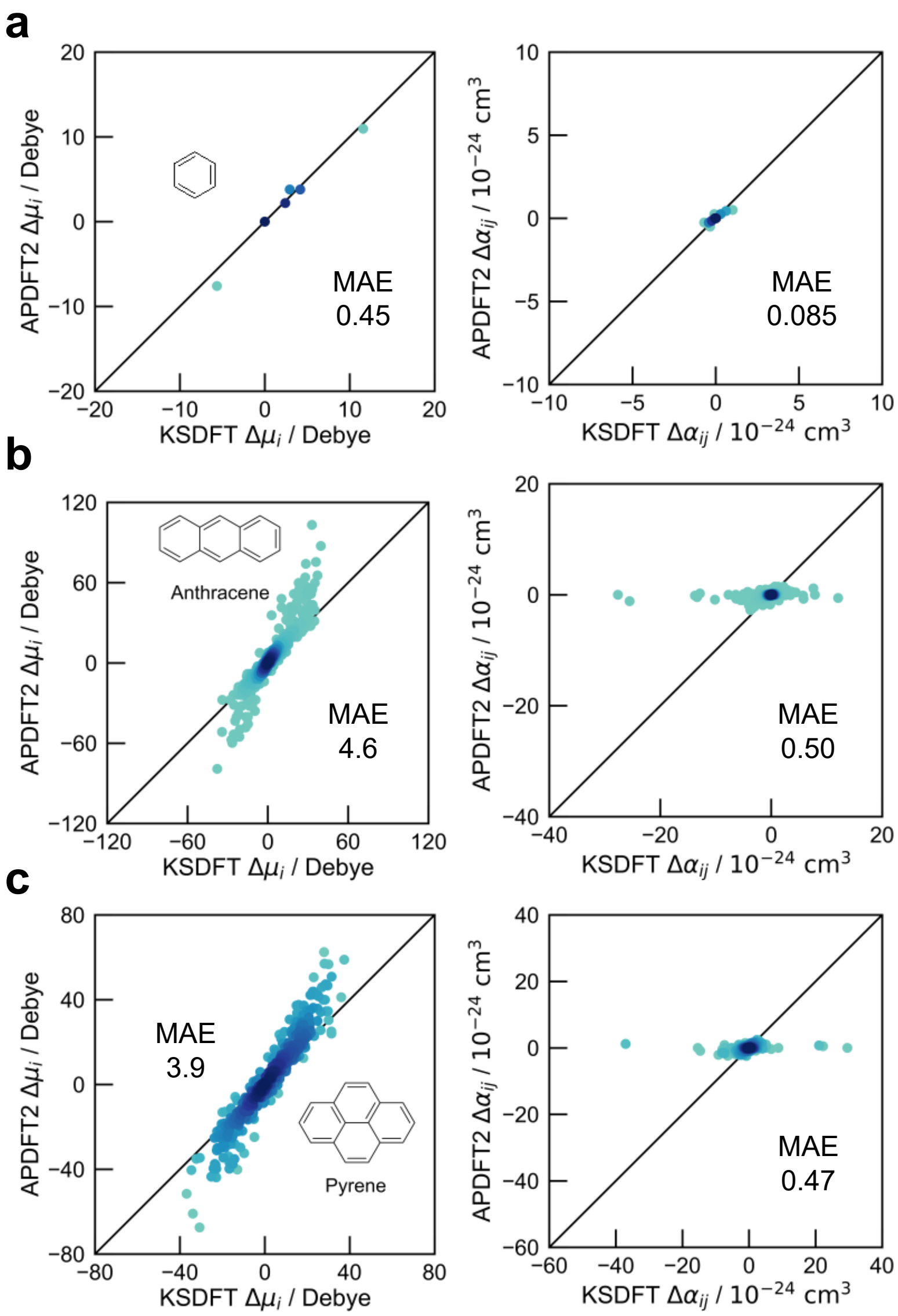

FIG. S10. APDFT2 predictions of the relative electric dipole moments ($\boldsymbol{\mu}$) and polarizabilities ($\boldsymbol{\alpha}$) of alchemical enantiomers and diastereomers for BN-doped (a) benzene, (b) anthracene, and (c) pyrene derivatives compared with the results of KSDFT. The intensity of the color represents the base-10 logarithm density. *i* and *j* are one of the axes of the Cartesian coordinates.

The locally stable structures of the pristine PAHs are shown here.

Benzene:

| | | | |
|---|---|---|---|
| C | 0.00000000 | 1.38730200 | 0.00000000 |
| C | 1.20143800 | 0.69365100 | 0.00000000 |
| C | 1.20143800 | -0.69365100 | 0.00000000 |
| C | 0.00000000 | -1.38730200 | 0.00000000 |
| C | -1.20143800 | -0.69365100 | 0.00000000 |
| C | -1.20143800 | 0.69365100 | 0.00000000 |
| H | 0.00000000 | 2.47054300 | 0.00000000 |
| H | 2.13955300 | 1.23527100 | 0.00000000 |
| H | 2.13955300 | -1.23527100 | 0.00000000 |
| H | 0.00000000 | -2.47054300 | 0.00000000 |
| H | -2.13955300 | -1.23527100 | 0.00000000 |
| H | -2.13955300 | 1.23527100 | 0.00000000 |

Naphthalene:

| | | | |
|---|---|---|---|
| C | 0.00000000 | 2.41711700 | 0.70411200 |
| C | 0.00000000 | 1.23656700 | 1.39335500 |
| C | 0.00000000 | 0.00000000 | 0.71091500 |
| C | 0.00000000 | 0.00000000 | -0.71091500 |
| C | 0.00000000 | 1.23656700 | -1.39335500 |
| C | 0.00000000 | 2.41711700 | -0.70411200 |
| C | 0.00000000 | -1.23656700 | 1.39335500 |
| C | 0.00000000 | -1.23656700 | -1.39335500 |
| C | 0.00000000 | -2.41711700 | -0.70411200 |
| C | 0.00000000 | -2.41711700 | 0.70411200 |
| H | 0.00000000 | -1.23360700 | 2.47755700 |
| H | 0.00000000 | 3.35868400 | 1.23943900 |
| H | 0.00000000 | 1.23360700 | 2.47755700 |
| H | 0.00000000 | 1.23360700 | -2.47755700 |
| H | 0.00000000 | 3.35868400 | -1.23943900 |
| H | 0.00000000 | -1.23360700 | -2.47755700 |
| H | 0.00000000 | -3.35868400 | -1.23943900 |
| H | 0.00000000 | -3.35868400 | 1.23943900 |

Anthracene:

| | | | |
|---|---|---|---|
| C | 0.00000000 | 3.63549700 | 0.70895200 |
| C | 0.00000000 | 2.46281500 | 1.39797800 |
| C | 0.00000000 | 1.21508800 | 0.71681300 |
| C | 0.00000000 | 1.21508800 | -0.71681300 |
| C | 0.00000000 | 2.46281500 | -1.39797800 |
| C | 0.00000000 | 3.63549700 | -0.70895200 |
| C | 0.00000000 | 0.00000000 | 1.39450000 |
| C | 0.00000000 | 0.00000000 | -1.39450000 |
| C | 0.00000000 | -1.21508800 | -0.71681300 |
| C | 0.00000000 | -1.21508800 | 0.71681300 |
| C | 0.00000000 | -2.46281500 | 1.39797800 |
| C | 0.00000000 | -3.63549700 | 0.70895200 |
| C | 0.00000000 | -3.63549700 | -0.70895200 |
| C | 0.00000000 | -2.46281500 | -1.39797800 |
| H | 0.00000000 | -2.46079400 | 2.48206400 |
| H | 0.00000000 | 0.00000000 | 2.47957900 |
| H | 0.00000000 | 4.57882000 | 1.24107800 |
| H | 0.00000000 | 2.46079400 | 2.48206400 |
| H | 0.00000000 | 2.46079400 | -2.48206400 |
| H | 0.00000000 | 4.57882000 | -1.24107800 |
| H | 0.00000000 | 0.00000000 | -2.47957900 |
| H | 0.00000000 | -4.57882000 | 1.24107800 |
| H | 0.00000000 | -4.57882000 | -1.24107800 |
| H | 0.00000000 | -2.46079400 | -2.48206400 |

Pyrene:

| | | | |
|---|---|---|---|
| C | 0.00000000 | 0.00000000 | 3.50012200 |
| C | 0.00000000 | 1.20273200 | 2.81354400 |
| C | 0.00000000 | 1.22697100 | 1.41881100 |
| C | 0.00000000 | 0.00000000 | 0.70881900 |
| C | 0.00000000 | -1.22697100 | 1.41881100 |
| C | 0.00000000 | -1.20273200 | 2.81354400 |
| C | 0.00000000 | 2.44816800 | 0.67602500 |
| C | 0.00000000 | 0.00000000 | -0.70881900 |
| C | 0.00000000 | 1.22697100 | -1.41881100 |
| C | 0.00000000 | 2.44816800 | -0.67602500 |
| C | 0.00000000 | 1.20273200 | -2.81354400 |
| C | 0.00000000 | 0.00000000 | -3.50012200 |
| C | 0.00000000 | -1.20273200 | -2.81354400 |
| C | 0.00000000 | -1.22697100 | -1.41881100 |
| C | 0.00000000 | -2.44816800 | -0.67602500 |
| C | 0.00000000 | -2.44816800 | 0.67602500 |
| H | 0.00000000 | 2.14019900 | -3.35766200 |
| H | 0.00000000 | -3.38326500 | 1.22445900 |
| H | 0.00000000 | -3.38326500 | -1.22445900 |
| H | 0.00000000 | 3.38326500 | 1.22445900 |
| H | 0.00000000 | 0.00000000 | 4.58323900 |
| H | 0.00000000 | 2.14019900 | 3.35766200 |
| H | 0.00000000 | -2.14019900 | 3.35766200 |
| H | 0.00000000 | 3.38326500 | -1.22445900 |
| H | 0.00000000 | 0.00000000 | -4.58323900 |
| H | 0.00000000 | -2.14019900 | -3.35766200 |

**Spatial symmetry of electron density derivatives with respect to static and uniform external fields**

We consider the spatial symmetry of electron density derivatives, $\partial^n \rho(\mathbf{r},\mathbf{F}) / \partial \mathbf{F}^n \big|_{\mathbf{F}=\mathbf{0}}$, which are required to describe response properties. $\rho(\mathbf{r},\mathbf{F})$ is electron density, $\rho(\mathbf{r})$, in the presence of a static and uniform external field. **F** is a field magnitude vector. Here, we examine inversion and reflection as spatial symmetry operations.

**Inversion.** We chose the origin of a Cartesian reference frame to correspond to the center of symmetry of a compound. Reference compounds of alchemical enantiomers are highly symmetric,[1] and its $\rho(\mathbf{r})$ may be totally symmetric under inversion. Therefore,

$$\rho(\mathbf{r}) = \rho(-\mathbf{r})$$

For $\rho(\mathbf{r},\mathbf{F})$,

$$\rho(\mathbf{r},\mathbf{F}) = \rho(-\mathbf{r},-\mathbf{F})$$

Clearly,

$$\rho(\mathbf{r},\mathbf{0}) = \rho(-\mathbf{r},\mathbf{0})$$

Using the above relations, the first-order derivative is written as

$$\begin{aligned}
\frac{\partial \rho}{\partial F_i}(-\mathbf{r},-\mathbf{F}) &= \lim_{h\to 0} \frac{\rho(-\mathbf{r},-F_i+h,-F_j,-F_k) - \rho(-\mathbf{r},-F_i,-F_j,-F_k)}{h} \\
&= -\lim_{h\to 0} \frac{\rho(-\mathbf{r},-F_i,-F_j,-F_k) - \rho(-\mathbf{r},-F_i+h,-F_j,-F_k)}{h} \\
&= -\lim_{h\to 0} \frac{\rho(\mathbf{r},F_i,F_j,F_k) - \rho(\mathbf{r},F_i-h,F_j,F_k)}{h} \\
&= -\frac{\partial \rho}{\partial F_i}(\mathbf{r},\mathbf{F})
\end{aligned}$$

where *i*, *j*, and *k* are the axes of the Cartesian coordinates. Therefore,

$$\frac{\partial \rho}{\partial \mathbf{F}}(\mathbf{r},\mathbf{F}) = -\frac{\partial \rho}{\partial \mathbf{F}}(-\mathbf{r},-\mathbf{F})$$

$\partial \rho(\mathbf{r},\mathbf{F}) / \partial \mathbf{F}$ is an odd function with respect to a pair of **r** and $\mathbf{F}$. At zero field strength, we obtain

$$\frac{\partial \rho}{\partial \mathbf{F}}(\mathbf{r},\mathbf{0}) = -\frac{\partial \rho}{\partial \mathbf{F}}(-\mathbf{r},\mathbf{0})$$

This equation shows that the first-order derivative is an odd function with respect to **r** when **F** = **0**. Using the above equation, we show that $\partial^2 \rho(\mathbf{r},\mathbf{F}) / \partial \mathbf{F}^2$ is an even function with respect to a pair of **r** and **F**:

$$\frac{\partial^2 \rho}{\partial {F_i}^2}(-\mathbf{r},-\mathbf{F}) = \lim_{h\to 0}\frac{\frac{\partial \rho}{\partial F_i}(-\mathbf{r},-F_i+h,-F_j,-F_k)-\frac{\partial \rho}{\partial F_i}(-\mathbf{r},-F_i,-F_j,-F_k)}{h}$$

$$= \lim_{h\to 0}\frac{-\frac{\partial \rho}{\partial F_i}(-\mathbf{r},-F_i,-F_j,-F_k)+\frac{\partial \rho}{\partial F_i}(-\mathbf{r},-F_i+h,-F_j,-F_k)}{h}$$

$$= \lim_{h\to 0}\frac{\frac{\partial \rho}{\partial F_i}(\mathbf{r},F_i,F_j,F_k)-\frac{\partial \rho}{\partial F_i}(\mathbf{r},F_i-h,F_j,F_k)}{h}$$

$$= \frac{\partial^2 \rho}{\partial {F_i}^2}(\mathbf{r},\mathbf{F})$$

and

$$\frac{\partial^2 \rho}{\partial F_i \partial F_j}(-\mathbf{r},-\mathbf{F}) = \lim_{h\to 0}\frac{\frac{\partial \rho}{\partial F_i}(-\mathbf{r},-F_i,-F_j+h,-F_k)-\frac{\partial \rho}{\partial F_i}(-\mathbf{r},-F_i,-F_j,-F_k)}{h}$$

$$= \lim_{h\to 0}\frac{-\frac{\partial \rho}{\partial F_i}(-\mathbf{r},-F_i,-F_j,-F_k)+\frac{\partial \rho}{\partial F_i}(-\mathbf{r},-F_i,-F_j+h,-F_k)}{h}$$

$$= \lim_{h\to 0}\frac{\frac{\partial \rho}{\partial F_i}(\mathbf{r},F_i,F_j,F_k)-\frac{\partial \rho}{\partial F_i}(\mathbf{r},F_i,F_j-h,F_k)}{h}$$

$$= \frac{\partial^2 \rho}{\partial F_i \partial F_j}(\mathbf{r},\mathbf{F})$$

We obtain

$$\frac{\partial^2 \rho}{\partial \mathbf{F}^2}(\mathbf{r},\mathbf{F}) = \frac{\partial^2 \rho}{\partial \mathbf{F}^2}(-\mathbf{r},-\mathbf{F})$$

When **F** = **0**,

$$\frac{\partial^2 \rho}{\partial \mathbf{F}^2}(\mathbf{r},\mathbf{0}) = \frac{\partial^2 \rho}{\partial \mathbf{F}^2}(-\mathbf{r},\mathbf{0})$$

Repeating the above procedure for the higher-order derivatives, it is generally shown that $\partial^n \rho(\mathbf{r},\mathbf{F}) / \partial \mathbf{F}^n \big|_{\mathbf{F}=\mathbf{0}}$ is odd and even with respect to **r** for odd and even *n*, respectively.

**Reflection.** We consider a reference compound with a mirror symmetry. Assuming that the reflection plane is located on the *jk* plane,

$$\rho(r_i,r_j,r_k) = \rho(-r_i,r_j,r_k)$$

For $\rho(\mathbf{r},\mathbf{F})$,

$$\rho\left(r_i,r_j,r_k,F_i,F_j,F_k\right)=\rho\left(-r_i,r_j,r_k,-F_i,F_j,F_k\right)$$

Obviously,

$$\rho\left(r_i,r_j,r_k,0,0,0\right)=\rho\left(-r_i,r_j,r_k,0,0,0\right)$$

Using the above relations, the first-order derivatives with respect to $F_i$ and $F_j$ are written as

$$\begin{aligned}
\frac{\partial\rho}{\partial F_i}\left(-r_i,r_j,r_k,-F_i,F_j,F_k\right)&=\lim_{h\to0}\frac{\rho\left(-r_i,r_j,r_k,-F_i+h,F_j,F_k\right)-\rho\left(-r_i,r_j,r_k,-F_i,F_j,F_k\right)}{h}\\
&=-\lim_{h\to0}\frac{\rho\left(-r_i,r_j,r_k,-F_i,F_j,F_k\right)-\rho\left(-r_i,r_j,r_k,-F_i+h,F_j,F_k\right)}{h}\\
&=-\lim_{h\to0}\frac{\rho\left(r_i,r_j,r_k,F_i,F_j,F_k\right)-\rho\left(r_i,r_j,r_k,F_i-h,F_j,F_k\right)}{h}\\
&=-\frac{\partial\rho}{\partial F_i}\left(r_i,r_j,r_k,F_i,F_j,F_k\right)
\end{aligned}$$

and

$$\begin{aligned}
\frac{\partial\rho}{\partial F_j}\left(-r_i,r_j,r_k,-F_i,F_j,F_k\right)&=\lim_{h\to0}\frac{\rho\left(-r_i,r_j,r_k,-F_i,F_j+h,F_k\right)-\rho\left(-r_i,r_j,r_k,-F_i,F_j,F_k\right)}{h}\\
&=\lim_{h\to0}\frac{\rho\left(r_i,r_j,r_k,F_i,F_j+h,F_k\right)-\rho\left(r_i,r_j,r_k,F_i,F_j,F_k\right)}{h}\\
&=\frac{\partial\rho}{\partial F_j}\left(r_i,r_j,r_k,F_i,F_j,F_k\right)
\end{aligned}$$

Further derivative with respect to $F_i$ results in

$$\begin{aligned}
\frac{\partial^2\rho}{\partial F_i^2}\left(-r_i,r_j,r_k,-F_i,F_j,F_k\right)&=\lim_{h\to0}\frac{\frac{\partial\rho}{\partial F_i}\left(-r_i,r_j,r_k,-F_i+h,F_j,F_k\right)-\frac{\partial\rho}{\partial F_i}\left(-r_i,r_j,r_k,-F_i,F_j,F_k\right)}{h}\\
&=\lim_{h\to0}\frac{-\frac{\partial\rho}{\partial F_i}\left(-r_i,r_j,r_k,-F_i,F_j,F_k\right)+\frac{\partial\rho}{\partial F_i}\left(-r_i,r_j,r_k,-F_i+h,F_j,F_k\right)}{h}\\
&=\lim_{h\to0}\frac{\frac{\partial\rho}{\partial F_i}\left(r_i,r_j,r_k,F_i,F_j,F_k\right)-\frac{\partial\rho}{\partial F_i}\left(r_i,r_j,r_k,F_i-h,F_j,F_k\right)}{h}\\
&=\frac{\partial^2\rho}{\partial F_i^2}\left(r_i,r_j,r_k,F_i,F_j,F_k\right)
\end{aligned}$$

and

$$
\begin{aligned}
\frac{\partial^2 \rho}{\partial F_i \partial F_j}\left(-r_i, r_j, r_k, -F_i, F_j, F_k\right) &= \lim_{h \to 0} \frac{\dfrac{\partial \rho}{\partial F_j}\left(-r_i, r_j, r_k, -F_i + h, F_j, F_k\right) - \dfrac{\partial \rho}{\partial F_j}\left(-r_i, r_j, r_k, -F_i, F_j, F_k\right)}{h} \\
&= -\lim_{h \to 0} \frac{\dfrac{\partial \rho}{\partial F_j}\left(-r_i, r_j, r_k, -F_i, F_j, F_k\right) - \dfrac{\partial \rho}{\partial F_j}\left(-r_i, r_j, r_k, -F_i + h, F_j, F_k\right)}{h} \\
&= -\lim_{h \to 0} \frac{\dfrac{\partial \rho}{\partial F_j}\left(r_i, r_j, r_k, F_i, F_j, F_k\right) - \dfrac{\partial \rho}{\partial F_j}\left(r_i, r_j, r_k, F_i - h, F_j, F_k\right)}{h} \\
&= -\frac{\partial^2 \rho}{\partial F_i \partial F_j}\left(r_i, r_j, r_k, F_i, F_j, F_k\right)
\end{aligned}
$$

As such, the derivative with respect to $F_i$ changes the sign, but the derivative with respect $F_j$ and $F_k$.do not. Therefore,

$$
\frac{\partial^{n_i+n_j+n_k} \rho}{\partial F_i^{n_i} \partial F_j^{n_j} \partial F_k^{n_k}}\left(r_i, r_j, r_k, F_i, F_j, F_k\right) = (-1)^{n_i} \frac{\partial^{n_i+n_j+n_k} \rho}{\partial F_i^{n_i} \partial F_j^{n_j} \partial F_k^{n_k}}\left(-r_i, r_j, r_k, -F_i, F_j, F_k\right)
$$

At zero field strength,

$$
\frac{\partial^{n_i+n_j+n_k} \rho}{\partial F_i^{n_i} \partial F_j^{n_j} \partial F_k^{n_k}}\left(r_i, r_j, r_k, 0, 0, 0\right) = (-1)^{n_i} \frac{\partial^{n_i+n_j+n_k} \rho}{\partial F_i^{n_i} \partial F_j^{n_j} \partial F_k^{n_k}}\left(-r_i, r_j, r_k, 0, 0, 0\right)
$$

The even- and odd-order derivatives with respect to $F_i$ are the same or sign opposite at symmetrically equivalent sites, respectively.

**Derivation of electric dipole moments and polarizabilities in APDFT**

We consider electric dipole moments and polarizabilities in APDFT. Using the Hellmann–Feynman theorem, APDFT$n$ electronic energy of isoelectronic materials is written as[2]

$$
\begin{aligned}
E^{\text{ele}} &= E_{\text{Ref}}^{\text{ele}} + \sum_n \frac{1}{n!}\frac{\partial^n E_{\text{Ref}}^{\text{ele}}}{\partial \lambda^n} \\
&= E_{\text{Ref}}^{\text{ele}} + \sum_n \frac{1}{n!}\frac{\partial^{n-1}}{\partial \lambda^{n-1}} \sum_I \frac{\partial E_{\text{Ref}}^{\text{ele}}}{\partial Z_I}\Delta Z_I \\
&= E_{\text{Ref}}^{\text{ele}} - \int d\mathbf{r} \sum_I \frac{\Delta Z_I}{\left|\mathbf{r}-\mathbf{R}_I\right|} \sum_n \frac{1}{n!}\frac{\partial^{n-1}\rho_{\text{Ref}}\left(\mathbf{r}\right)}{\partial \lambda^{n-1}}
\end{aligned}
$$

where $\lambda$ ($0 \le \lambda \le 1, \lambda \in \mathbb{R}$) is the coupling parameter defined in the main text, $E_{\text{Ref}}^{\text{ele}}$ is electronic energy of the reference material, $Z_I$ is a nuclear charge of the $I$th atom of the reference material, $\Delta Z_I$ is the change in the nuclear charge from the reference material to a different material, and $\rho_{\text{Ref}}\left(\mathbf{r}\right)$ is electron density of the reference material. The electronic part of the electric dipole moment is the first-order derivative of $E_{\text{Ref}}^{\text{ele}}$ with respect to an electric field amplitude vector **E**:

$$
\boldsymbol{\mu}^{\mathbf{ele}} = -\left.\frac{\partial E^{\text{ele}}\left(\mathbf{E}\right)}{\partial \mathbf{E}}\right|_{\mathbf{E}=\mathbf{0}}
$$

Substituting the APDFT1 electronic energy results in[2, 3]

$$
\begin{aligned}
\boldsymbol{\mu}^{\mathbf{ele}} &\approx -\left.\frac{\partial E_{\text{Ref}}^{\text{ele}}\left(\mathbf{E}\right)}{\partial \mathbf{E}}\right|_{\mathbf{E}=\mathbf{0}} - \sum_I \left.\frac{\partial^2 E_{\text{Ref}}^{\text{ele}}\left(\mathbf{E}\right)}{\partial Z_I \partial \mathbf{E}}\right|_{\mathbf{E}=\mathbf{0}} \Delta Z_I \\
&= \boldsymbol{\mu}_{\text{Ref}}^{\text{ele}} + \int d\mathbf{r} \sum_I \frac{\Delta Z_I}{\left|\mathbf{r}-\mathbf{R}_I\right|} \left.\frac{\partial \rho_{\text{Ref}}\left(\mathbf{r},\mathbf{E}\right)}{\partial \mathbf{E}}\right|_{\mathbf{E}=\mathbf{0}} \\
&= \boldsymbol{\mu}_{\text{Ref}}^{\text{ele}} - \int d\mathbf{r}\, \mathbf{r} \sum_I \left.\frac{\partial \rho_{\text{Ref}}\left(\mathbf{r},\mathbf{E}\right)}{\partial Z_I}\right|_{\mathbf{E}=\mathbf{0}} \Delta Z_I
\end{aligned}
$$

where $\boldsymbol{\mu}_{\text{Ref}}^{\text{ele}}$ is the electric dipole moment of the reference material. The last two equations give the same results because of the Schwartz theorem but require different types of orbital rotation matrices for $Z_I$ and **E**. Within APDFT1, the electric polarizability is written as

$$
\begin{aligned}
\boldsymbol{\alpha}^{\mathbf{ele}} &= -\left.\frac{\partial^2 E^{\text{ele}}\left(\mathbf{E}\right)}{\partial \mathbf{E}^2}\right|_{\mathbf{E}=\mathbf{0}} \\
&\approx \boldsymbol{\alpha}_{\text{Ref}}^{\text{ele}} - \sum_I \left.\frac{\partial^3 E_{\text{Ref}}^{\text{ele}}\left(\mathbf{E}\right)}{\partial Z_I \partial \mathbf{E}^2}\right|_{\mathbf{E}=\mathbf{0}} \Delta Z_I
\end{aligned}
$$

where $\boldsymbol{\alpha}_{\text{Ref}}^{\text{ele}}$ represents the electric polarizability of the reference material. We implemented analytical $\left.\partial^2 E_{\text{Ref}}^{\text{ele}}\left(\mathbf{E}\right)/\partial Z_I \partial \mathbf{E}\right|_{\mathbf{E}=\mathbf{0}}$ and $\left.\partial^3 E_{\text{Ref}}^{\text{ele}}\left(\mathbf{E}\right)/\partial Z_I \partial \mathbf{E}^2\right|_{\mathbf{E}=\mathbf{0}}$ in KSDFT with Pulay's equations.[4] The coupled perturbed Kohn–Sham equations are solved to compute orbital rotation matrices for $Z_I$ and **E**.

## References in the supplementary material